\pdfoutput=1

\documentclass[a4paper,twocolumn]{article}

\usepackage{subcaption}
\usepackage{bm}

\usepackage{mathtools}
\DeclarePairedDelimiter{\norm}{\lVert}{\rVert}
\DeclarePairedDelimiter\abs{\lvert}{\rvert}

\usepackage{amsmath}
\usepackage{amsfonts}

\usepackage{siunitx}
\DeclareSIUnit\angstrom{\text {Å}}
\usepackage{float}
\usepackage{graphicx} 

\usepackage{hyperref}

\hypersetup{
    pdftitle={X-ray dark-field via spectral propagation-based imaging},    
    pdfauthor={Jannis Ahlers},     
    pdfsubject={X-ray dark-field via spectral propagation-based imaging},   
    pdfcreator={Jannis Ahlers},   
    pdfproducer={Jannis Ahlers}, 
    pdfkeywords={X-ray} {Imaging} {Dark-field} {Spectral} {Dual-energy} {Propagation-based imaging} {PBI}, 
}

\usepackage{cleveref}
\creflabelformat{equation}{#2\textup{#1}#3}

\usepackage[margin={2cm,2cm}]{geometry}
\title{X-ray dark-field via spectral propagation-based imaging}
\author{Jannis N. Ahlers$^{1,*}$,~Konstantin M. Pavlov$^{2,1,3}$,~Marcus J. Kitchen$^{1}$,~and Kaye S. Morgan$^{1}$}
\date{\small{$^{1}$ School of Physics and Astronomy, Monash University, Clayton VIC 3800, Australia\\
$^{2}$ School of Physical and Chemical Sciences, University of Canterbury, Christchurch 8140, New Zealand\\
$^{3}$ School of Science and Technology, University of New England, Armidale NSW 2351, Australia\\
$^{*}$ Corresponding author: jannis.ahlers@monash.edu}}

\begin{document}

\twocolumn[
  \begin{@twocolumnfalse}
    \maketitle
    \begin{abstract}
        \noindent
        Dark-field X-ray imaging is a novel modality which visualises scattering from unresolved microstructure. Most dark-field imaging techniques rely on crystals or structured illumination, but recent work has shown that dark-field effects are observable in straightforward propagation-based imaging (PBI). Based on the single-material X-ray Fokker--Planck equation with an a priori dark-field energy dependence, we propose an algorithm to extract phase and dark-field effects from dual-energy PBI images. We successfully apply the dark-field retrieval algorithm to simulated and experimental dual-energy data, and show that by accounting for dark-field effects, projected thickness reconstruction is improved compared to the classic Paganin algorithm. With the emergence of spectral detectors, the method could enable single-exposure dark-field imaging of dynamic and living samples.
    \end{abstract}
    \bigskip
  \end{@twocolumnfalse}
]


\section{Introduction}
X-ray imaging is a cheap, fast, accessible, high-resolution imaging technique. It is ubiquitous in medical, security, and industrial applications; together, planar and computed-tomography (CT) X-ray scans make up more than 50\% of diagnostic medical imaging \cite{smith-bindman2012}. However, traditional attenuation-based X-ray imaging has two major drawbacks compared to competing modalities, such as magnetic resonance imaging (MRI) and ultrasound: it comes at a cost of absorbed dose from the ionizing X-rays, and it provides poor contrast in weakly attenuating soft tissues. The advent of high-coherence synchrotron and micro-focus X-ray sources has led to the development of phase-contrast and dark-field X-ray imaging modalities, which supplement attenuation-based contrast \cite{paganin2006}. Phase-contrast X-ray imaging measures the refraction of X-rays in the sample, and can achieve significantly improved image quality compared to traditional X-ray imaging, particularly in soft tissue \cite{kitchen2017}. Dark-field X-ray imaging visualises the scattered component of the X-ray beam \cite{zhong2000, pfeiffer2008}.
It is associated with small-angle X-ray scattering (SAXS) from unresolved microstructure in the sample, and is a promising candidate for the investigation of such structure. 
Examples of its use include studying water transport in porous media \cite{yang2014},
diagnosis and assessment of pulmonary diseases \cite{yaroshenko2015, kitchen2020, meinel2014a} and breast cancer \cite{wang2014}, 
and detection of crystallised urate depositions \cite{braig2020}.

A number of techniques have been developed for dark-field imaging. The earliest X-ray dark-field imaging reflected the X-ray wavefield from a rotating crystal to analyse the spread in propagation direction, a technique known as analyser-based or diffraction-enhanced imaging \cite{pagot2003, rigon2007, kitchen2010}. More recent methods imprint a known reference pattern (structured-illumination/coded aperture) onto the X-ray wavefront, which is then modulated by the presence of the sample, causing changes in amplitude, position, and visibility of the reference pattern (associated respectively with the attenuation, phase-shift, and scattering of the X-rays). If the pixel size is larger than the reference pattern structure, a scanning step is generally used to fully recover the modulation, for example in Talbot--Lau interferometry \cite{david2002, momose2003, pfeiffer2008} and edge-illumination \cite{endrizzi2014, olivo2021}. If the reference pattern is well resolved, modulations can be imaged directly without the need for phase stepping, such as in single-grid imaging \cite{wen2010, morgan2011}, speckle-based imaging \cite{berujon2012, morgan2012}, and beam-tracking edge-illumination \cite{dreier2020}.

In these latter direct-imaging techniques, methods of reconstructing phase-shift and dark-field generally fall into `local' or `global' camps, sometimes referred to as `explicit' and `implicit' tracking. Local methods rely on the modulation pattern having `small' (high-frequency) features, usually of order a few pixels; a window around each pixel is applied, and the modulation within that window is explicitly calculated. Examples of this are cross-correlation analysis \cite{morgan2011, berujon2012a} and unified modulated pattern analysis (UMPA) \cite{zdora2017} in single-grid and speckle-based imaging. 
An alternative approach is to model the changes in the wavefront globally, and solve the inverse form of this model for the entire image at once. An example of this is the use of geometric flow \cite{paganin2018} and the transport-of-intensity equation (TIE) \cite{pavlov2020a} in speckle-based imaging. The TIE models the coherent flow of optical energy in a propagating paraxial complex-scalar wavefield \cite{teague1983}. An extension of the TIE that includes diffusive flow is the Fokker--Planck equation of X-ray imaging \cite{paganin2019, morgan2019}. For a $z$-paraxial plane wave propagating forward after passing through a thin object, the equation can be given in its finite-difference form for the near-field regime as    
\begin{equation}
\label{eq:FP}
\begin{split}
    I(x, y, z = \Delta) &= I(x,y,z=0) \\ &- \frac{\Delta}{k}\nabla_\perp \cdot [I(x,y,z)\nabla_\perp \phi(x,y,z)]_{z=0} \\ &+ \Delta^2 \nabla_\perp^2[D(x,y) I(x,y,z)]_{z=0},
\end{split}
\end{equation}
where the exit-surface of the sample is at $z=0$, $I(x,y,z)$ is the intensity of the wavefield, $\Delta$ is the propagation distance from the exit-surface of the sample to the detector, $k = 2\pi / \lambda$ is the wavenumber, $\lambda$ is the wavelength, $\nabla_\perp \equiv (\partial_x, \partial_y)$ is the transverse gradient operator, $\phi(x,y,z)$ is the phase-shift of the wavefield caused by the object, and $D(x,y)$ is the dimensionless X-ray Fokker--Planck diffusion coefficient, which is related to the angular spread of the dark-field. Excluding the diffusion term from \cref{eq:FP} recovers the finite-difference form of the TIE. The X-ray Fokker--Planck equation can be used to model grating-based imaging \cite{morgan2019}, and has been used as the basis for novel dark-field retrieval methods in speckle-based imaging \cite{pavlov2020, alloo2022}.

A particularly simple X-ray phase-contrast and dark-field imaging technique, which does not require patterning of the wavefront, is propagation-based imaging (PBI). In PBI, an unpatterned spatially-coherent wavefront interacts with the sample, and is subsequently propagated through free space. Phase shifts introduced by the sample cause interference fringes to be developed in the free-space propagating intensity, which are then imaged by a detector \cite{snigirev1995, cloetens1996}. PBI does not require specialised optics or a monochromatic source, at the expense of strict spatial coherence requirements \cite{wilkins1996}. Single-image TIE-based phase-retrieval under a single-material assumption \cite{paganin2002} has been extensively applied for phase-contrast PBI, with the single-material assumption not preventing widespread adoption for diverse samples \cite{paganin2020}. PBI has recently been extended to dark-field imaging \cite{gureyev2020, aminzadeh2022, leatham2023}. Leatham~\textit{et al.}~\cite{leatham2023} developed a theory of PBI dark-field imaging of a single-material object, based on the Fokker--Planck equation. The method requires imaging a sample at two different propagation distances, relying on the geometric dependence of the scattering cone to separate dark-field effects from attenuation and refraction of the beam. They noted that an alternative approach may be to change the energy of the beam. A multi-energy approach would not require changes in the experimental geometry while imaging, improving ease of implementation and reducing set-up stability requirements. In addition, the recent developments in energy-resolving detectors mean such an approach could capture all the required information within a single exposure.

Dual-energy imaging is a well established technique dating back to the 1980's \cite{speller1983}. A number of technologies for dual-energy imaging have been developed, such as rapid voltage switching, dual-source CT, and layered detectors \cite{alkadhi2022}. As the attenuation by a specific material is strongly energy-dependent with a theoretically well-grounded dependence, dual-energy imaging establishes a basis for material weighting or decomposition \cite{alvarez1976}. There are numerous clinical applications of dual-energy imaging, such as measuring bone mineral density, bone removal, and virtual non-contrast imaging \cite{alkadhi2022}. Dual- and multi-energy images have also been utilised in propagation-based phase-contrast imaging, using the TIE as a basis for phase-retrieval \cite{gureyev1998a,gureyev2001} and material decomposition \cite{schaff2020,schaff2022}. As dark-field is related to scattering, it has a strong dependence on X-ray energy. This dependence has been exploited in grating-based imaging to recover information about size of the sample microstructure \cite{taphorn2020a, partridge2022, taphorn2023}, to do material decomposition \cite{sellerer2021}, and to improve the signal-to-noise ratio of the recovered dark-field signal \cite{pelzer2014}.

This paper demonstrates a proof-of-concept for spectral propagation-based dark-field imaging (SPB-DF), based on a single-material Fokker--Planck model. Using dual-energy PBI images, the method initially reconstructs the sample's projected thickness. As a second step, the dark-field image is reconstructed by solving the diffusion term in the Fokker--Planck equation for the diffusion coefficient. In addition to using a classic numerical Poisson solver for this step, we introduce a method inspired by structured-illumination dark-field imaging techniques that measures a local change of visibility of the sample, exploiting the texture created by the sample as a `self-reference' pattern. SPB-DF enables optics-free dark-field imaging, without the necessity to move the sample. The potential use of an energy-discriminating detector and polychromatic source with SPB-DF would enable static, single-shot dark-field imaging, with promising applications for imaging dynamic processes.            

\section{Theory}
\label{sec:theory}

Under the Fokker--Planck model, dark-field effects are quantified in the diffusion coefficient $D$ (see \cref{eq:FP}). Alternatively, a measure of dark-field is the change in visibility $V$ of a reference pattern induced by dark-field blurring. Visibility in a small region-of-interest around each pixel can be measured as \cite{michelson1927,zdora2018} 
\begin{equation}
\label{eq:vis}
    V = \frac{I_\text{max} - I_\text{min}}{I_\text{max} + I_\text{min}} \approx \frac{\text{StdDev(I)}}{\text{Mean(I)}}.
\end{equation}
If the reference pattern can locally be modelled as a sinusoidal variation with period $p$, the visibilities of reference (i.e. no dark-field, $V_\text{ref}$) and observed (dark-field present, $V_\text{obs}$) images can be related to the Fokker--Planck diffusion coefficient (eq.~132 in \cite{paganin2023}, in the case where the dark-field coefficient does not vary significantly over a grid period \cite{morgan2019}) via
\begin{equation}
\label{eq:D_from_vis}
    \frac{V_\text{obs}}{V_\text{ref}} = \exp\left( \frac{-4\pi^2 \Delta^2 D}{p^2} \right).
\end{equation} 
Both the size of the window used in the measurement of visibility, and the parameter $p$ used in the conversion to $D$, depend on the local length scale of features in the reference image. \Cref{eq:D_from_vis} was first derived in the context of speckle and single-grid dark-field imaging, in which there is generally a well-defined length scale associated respectively with the mean speckle size and the grid period. In propagation-based imaging, the beam is not imprinted with a reference pattern. However, if the sample itself creates a relatively quickly-varying intensity, this can be treated as a self-reference pattern \cite{leatham2023}. For the purposes of this work, we will assume that the length scale of this self-reference pattern is constant throughout the image, and will measure it as twice the full-width at half-maximum of the central peak of a radially-averaged 2D autocorrelation -- this number is then used as both the window size and period $p$. This assumption is unlikely to hold for general samples imaged using propagation-based imaging, and in \cref{sec:discussion} we discuss possible methods that could be used in future work to make local estimates of a dominant length scale. In this paper we will reconstruct dark-field images by solving for the Fokker--Planck diffusion coefficient. We will use two approaches to do this, a `global' approach that numerically solves the diffusion part of the Fokker--Planck equation (a Poisson equation) for $D(x,y)$ directly, and a `local' approach that measures visibility in small regions using \cref{eq:vis} and converts it to $D(x,y)$ using \cref{eq:D_from_vis}.

When two images are taken at relatively low and high energy (cf. large and small propagation distance \cite{leatham2023}), dark-field diffusion is much stronger in the lower energy image. We aim to develop a theory for dark-field reconstruction using dual-energy images, based on the Fokker--Planck equation. To that end, consider a non-crystalline non-magnetic single-material sample being illuminated by a plane wavefield of unit intensity. Assuming a sample with projected thickness $T(x,y)$ and complex refractive index $n = 1 - \delta + i\beta$, with a propagation distance significantly larger than $T(x,y)$, and applying the projection approximation \cite{morgan2010}, the Fokker--Planck equation (\cref{eq:FP}) becomes
\begin{equation}
\label{eq:FPSM}
\begin{split}
    I(x, y, z = \Delta) &= e^{-\mu T(x,y)} -\frac{\Delta\delta }{\mu}\nabla_\perp^2 e^{-\mu T(x,y)} \\ &+ \Delta^2 \nabla_\perp^2\left[D(x,y) e^{- \mu T(x,y)}\right],
\end{split}
\end{equation}
where $\mu = 2k\beta$ is the linear attenuation coefficient. Our aim is to solve \cref{eq:FPSM} for the projected sample thickness $T(x,y)$ and the Fokker--Planck diffusion coefficient $D(x,y)$, by measuring the propagated intensity $I(x,y,z=\Delta)$ at two different energies. To solve for $T(x,y)$, we begin to linearise \cref{eq:FPSM} in $T(x,y)$ and $D(x,y)$ by assuming that $D(x,y)$ varies slowly enough that we can neglect higher order terms, giving $\nabla_\perp^2\left[D(x,y) e^{-\mu T(x,y)}\right] \approx D(x,y) \nabla_\perp^2 e^{-\mu T(x,y)}$. We expand the Laplacian (for brevity we drop $(x,y)$ arguments), giving
\begin{equation}
\label{eq:deriv}
    I \approx \left[ 1 + \left(\frac{\Delta \delta}{\mu} - \Delta^2 D\right)(\mu \nabla_\perp^2 T - \mu^2 \norm{\nabla_\perp T}^2)  \right] e^{-\mu T},
\end{equation}
where $\norm{\nabla_\perp T}^2 = \nabla_\perp T \cdot \nabla_\perp T$.
Let us assume that the sample is weakly-attenuating, with $\mu T \ll 1$. Then for most such samples $\mu^2 \norm{\nabla_\perp T}^2 \ll \abs{\mu \nabla_\perp^2 T}$, and we can neglect the smaller term. Then we multiply both sides of \cref{eq:deriv} by $1-\mu T$, and neglect higher order terms in $\mu T \nabla_\perp^2 T$, giving
\begin{align}
    I \frac{1-\mu T}{e^{-\mu T}} &\approx (1-\mu T)(1 + (\Delta \delta - \Delta^2 \mu D)\nabla_\perp^2 T) \\
    &\approx 1 - \mu T + \Delta \delta \nabla^2_\perp T - \Delta^2 \mu D \nabla^2_\perp T. \label{eq:FPSM-lin}
\end{align}
In first approximation we assume that $(1-\mu T)/(e^{-\mu T}) \approx 1$, with the aim of later iteratively correcting this approximation (\cref{eq:iter_1}). This gives us a partial-differential equation in two unknown functions $T(x,y)$ and $D(x,y)$, which we can solve using two measurements $I(x,y,z=\Delta)$ at two energies $E_1$ and $E_2$. We assume that, for our combination of (monochromatic) energies and sample, the Fokker--Planck diffusion coefficient can be decomposed as $D = D_0 \Gamma(\lambda)$, where $\lambda$ is the wavelength, $D_0$ is the energy-independent dark-field signal created by the sample, and $\Gamma(\lambda)$ is an unknown function describing the dark-field signal's dependence on energy. Inserting this definition into \cref{eq:FPSM-lin} and dividing by $\mu \Gamma(\lambda)$, we get
\begin{equation}
\label{eq:main}
    \frac{I}{\mu \Gamma(\lambda)} \approx \frac{1}{\mu \Gamma(\lambda)} - \frac{T}{\Gamma(\lambda)} + \frac{\Delta \delta}{\mu \Gamma(\lambda)} \nabla^2_\perp T - \Delta^2 D_0 \nabla^2_\perp T.
\end{equation}
The last term does not depend on energy. Therefore, for two intensity measurements at energies $E_1$ and $E_2$ (with corresponding wavelengths $\lambda_1$ and $\lambda_2$), we can write
\begin{equation}
    \begin{split}
        I_1 \mu_2 \Gamma(\lambda_2) &-  I_2 \mu_1 \Gamma(\lambda_1) =\\
        &\Delta(\delta_1 \mu_2 \Gamma(\lambda_2) - \delta_2 \mu_1 \Gamma(\lambda_1)) \nabla^2_\perp T \\
        &+ \mu_1 \mu_2 (\Gamma(\lambda_1) - \Gamma(\lambda_2)) T \\
        &+ (\mu_2 \Gamma(\lambda_2) - \mu_1 \Gamma(\lambda_1)).
    \end{split}
\end{equation}
Denote $\mathfrak{F}[f(x,y)] = \iint_{-\infty}^{\infty} f(x,y) \exp{(-2\pi i \vec{r} \cdot \vec{\xi})} \mathop{d\vec{r}}$ as the 2D Fourier transform, with the Fourier-space coordinates $\vec{\xi} = (\xi_x, \xi_y)$ dual to the real-space coordinates $\vec{r} = (x, y)$. We take the Fourier transform of both sides and apply the Fourier derivative theorem, leading to
\begin{equation}
    \begin{split}
        \mathfrak{F} &[ I_1 \mu_2 \Gamma(\lambda_2) -  I_2 \mu_1 \Gamma(\lambda_1) ] = \\
        &- 4 \pi^2 |\vec{\xi}|^2 \Delta(\delta_1 \mu_2 \Gamma(\lambda_2) - \delta_2 \mu_1 \Gamma(\lambda_1)) \mathfrak{F} [T] \\
        &+ \mu_1 \mu_2 (\Gamma(\lambda_1) - \Gamma(\lambda_2)) \mathfrak{F} [T] \\
        &+ (\mu_2 \Gamma(\lambda_2) - \mu_1 \Gamma(\lambda_1))\bm{\delta}(\vec{\xi}),
    \end{split}
\end{equation}
where $\bm{\delta}(\vec{\xi})$ is the 2D Dirac delta function. Finally, we rearrange for $\mathfrak{F} [T]$ and apply the inverse Fourier transform to get
\begin{equation}
\label{eq:proj_thick}
\begin{split}
    T &= \mathfrak{F}^{-1} \left [ \frac{\mathfrak{F} \left[ \mu_2 \Gamma(\lambda_2) I_1 - \mu_1 \Gamma(\lambda_1) I_2 \right]}
    {f(\lambda_1,\lambda_2,\vec{\xi})} \right ]\\ 
    &+ \frac{\mu_1 \Gamma(\lambda_1) - \mu_2 \Gamma(\lambda_2)}{\mu_1 \mu_2 (\Gamma(\lambda_1) - \Gamma(\lambda_2))}
\end{split}
\end{equation}
where
\begin{equation}
    \begin{split}
        f(\lambda_1,\lambda_2,\vec{\xi}) &= \mu_1 \mu_2 (\Gamma(\lambda_1) - \Gamma(\lambda_2)) \\ &- 4 \pi^2 |\vec{\xi}|^2 \Delta (\delta_1 \mu_2 \Gamma(\lambda_2) - \delta_2 \mu_1 \Gamma(\lambda_1)).
    \end{split}
\end{equation}
\Cref{eq:proj_thick} gives an initial estimate $T^{(0)}$ of the sample projected thickness, using two recorded images at two different energies. To improve the estimate, we iteratively correct the left-hand-side in the fully linearised Fokker--Planck equation (\cref{eq:FPSM-lin}) using
\begin{equation}
\label{eq:iter_1}
    I^{(\zeta + 1)} = I^{(\zeta)} \cdot \frac{1 - \mu T^{(\zeta)}}{e^{-\mu T^{(\zeta)}}},
\end{equation}
and re-solve for $T^{(\zeta + 1)}$, where $\zeta$ is the index of iteration. This iterative process can be carried out for a chosen number of iterations, or until a stability criterion is met (see \cref{sec:exp} and \cref{fig:iter} for further information).

After finding the projected thickness $T(x,y)$, the only unknown left in the single-material Fokker--Planck equation is the diffusion coefficient $D(x,y)$. With only the diffusion term, we are left with a two-dimensional Poisson equation. A variety of methods are available to numerically solve the discrete Poisson equation \cite{paganin2023}. Because it is straightforward and computationally inexpensive, we use a de-convolution in Fourier space (see eq.~13 in \cite{leatham2023}) given as
\begin{equation}
\label{eq:D_InvLap}
\begin{split}
    D &= \frac{e^{\mu T}}{\Delta^2} \nabla_\perp^{-2} \biggl[ I - \underbrace{\left(1 - \frac{\delta \Delta}{\mu} \nabla^2_\perp \right) e^{-\mu T}}_{I_{\text{DF-free}}} \biggr],
\end{split}
\end{equation}
where the inverse Laplacian $\nabla_\perp^{-2}$ is a pseudo-differential operator of the form
\begin{equation}
    \nabla_\perp^{-2} = -\mathfrak{F}^{-1} \frac{1}{4 \pi^2 |\vec{\xi}|^2 + \varepsilon} \mathfrak{F}.
\end{equation}
The singularity at the origin of the Fourier filter destabilises the reconstruction at low frequencies, and must be regularised, in this case using a Tikhonov regularisation parameter $\varepsilon$. This parameter is fine-tuned for each image to suppress low-frequency artefacts. 

We refer to using a numerical solution to the Poisson equation, in our case \cref{eq:D_InvLap}, as a `global' method of reconstructing dark-field. Alternatively, we could consider a `local' approach based on measuring visibility (\cref{eq:vis}) and converting to $D(x,y)$ using \cref{eq:D_from_vis}. \Cref{eq:D_from_vis} holds directly for pure-dark-field objects, but does not account for attenuation and phase effects. Our low energy image $I_{E_1}$ contains a strong dark-field signal, and can act as the `observed' image. A matching `reference' image needs to have identical attenuation and phase effects, but without the dark-field blurring present in $I_{E_1}$. We can create such an image using our reconstructed projected thickness $T(x,y)$ --- by simulating a phase-contrast image at the lower energy using a forward model that does \emph{not} include dark-field effects, a `virtual' dark-field-free image is created.
An example of such a forward model appears in the inverted Fokker--Planck equation (\cref{eq:D_InvLap}); the term inside the inverse Laplacian denoted $I_{\text{DF-free}}$ is the transport-of-intensity equation propagator, applied to a thin, single-material object \cite{paganin2006}. Up to approximation, $I$ and $I_{\text{DF-free}}$ should then be identical anywhere there is no dark-field, but differ proportionally with $\nabla^2_\perp (DI)$ where $D \neq 0$. The visibility maps $V_\text{obs}(x,y)$ and $V_\text{ref}(x,y)$ can therefore be calculated from $I$ and $I_{\text{DF-free}}$ respectively (using \cref{eq:vis}), and then inserted into \cref{eq:D_from_vis} to recover a dark-field image.

\section{Simulation}
\label{sec:sim}

\begin{figure}
    \centering
    \includegraphics[width=\columnwidth]{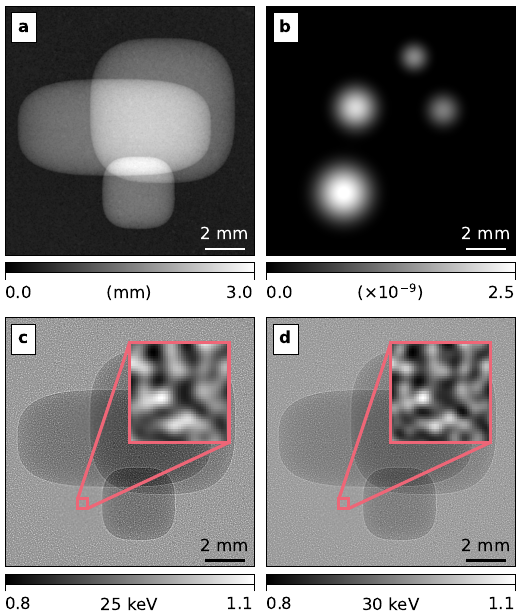}
    {\phantomsubcaption\label{fig:sim-simT}}
    {\phantomsubcaption\label{fig:sim-simD}}
    {\phantomsubcaption\label{fig:sim-I-E1}}
    {\phantomsubcaption\label{fig:sim-I-E2}}
    \caption{Simulated data. (\subref{fig:sim-simT})~The projected thickness map $T_\text{in}(x,y)$ consists of three superellipsoids, in addition to quickly-varying features that cover the entire image. (\subref{fig:sim-simD})~The simulated dark-field signal in the form of the Fokker--Planck diffusion coefficient $D(x,y)$ at \qty{25}{\keV}. The bottom row contains the propagated and dark-field-blurred intensities at (\subref{fig:sim-I-E1})~\qty{25}{\keV} and (\subref{fig:sim-I-E2})~\qty{30}{\keV}. To show the difference in local dark-field-associated blurring at the two energies, zoomed insets in panels (\subref{fig:sim-I-E1}) and (\subref{fig:sim-I-E2}) have been included, with their greyscales adjusted to the local minimum and maximum. While the difference in subtle, the reconstruction is able to extract this signal.}
    \label{fig:sim_sim-data}
\end{figure}

\begin{figure}
    \centering
    \includegraphics[width=\columnwidth]{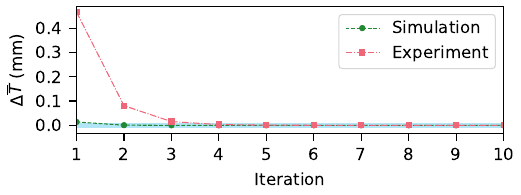}
    \caption{Difference in mean reconstructed projected thickness between each iteration $n$ of the iterative correction to our algorithm, where $\Delta \overline{T} = \overline{T^{(n)}} - \overline{T^{(n-1)}}$. The shaded area indicates \qty{\pm 10}{\micro\meter}. Both simulated (\cref{sec:sim}) and experimental data (\cref{sec:exp}) show rapid convergence.}
    \label{fig:iter}
\end{figure}

\begin{figure}
    \centering
    \includegraphics[width=\columnwidth]{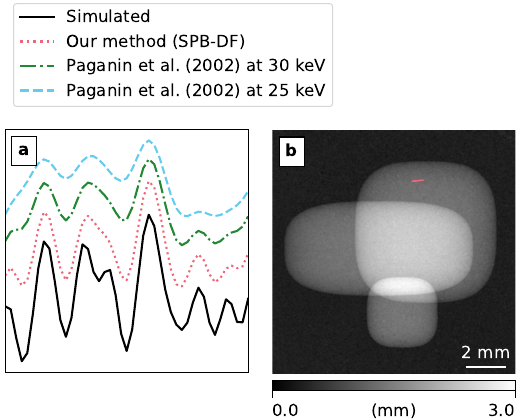}
    {\phantomsubcaption\label{fig:sim-T_lp}}
    {\phantomsubcaption\label{fig:sim-T}}
    \caption{(\subref{fig:sim-T_lp})~An intensity line profile comparison between (\subref{fig:sim-T})~reconstruction of the projected thickness $T(x,y)$ using SPB-DF, and single-material phase retrieval of the \qty{25}{\keV} and \qty{30}{\keV} images based on the transport-of-intensity equation \cite{paganin2002}. The solid line shows the simulated projected thickness. By accounting for the strong dark-field diffusion in this part of the sample, SPB-DF improves the fidelity of reconstruction.}
    \label{fig:sim_thick}
\end{figure}

An initial validation of the algorithm described in \cref{sec:theory} was carried out via simulation. A projected thickness map $T_\text{in}(x,y)$ was generated (\cref{fig:sim-simT}), consisting of the sum of a bulk thickness and a quickly-varying texture to simulate a realistic sample. The bulk thickness was created by overlaying three projected superellipsoids, and the quickly-varying component of the projected thickness was derived from an X-ray image of a random absorption mask \cite{quenot2021, labriet2021}. This pattern had a mean period of approximately ten pixels, measured as twice the full-width at half-maximum of the central peak of a radially-averaged 2D autocorrelation. The $T_\text{in}(x,y)$ map consisted of $1032 \times 1032$~pixels with a pixel size of \qty{12}{\micro\meter}.

Complex refractive index values $\delta$ and $\beta$ for poly(methyl methacrylate) (PMMA) at the two chosen energies $E_1=\qty{25}{\keV}$ and $E_2=\qty{30}{\keV}$ were found using \texttt{xraylib} \cite{schoonjans2011}. Under the projection approximation, the exit-surface complex scalar wavefield was calculated as
\begin{equation}
    \Psi(x,y,z=0) = e^{-kT_\text{in}(\beta + i\delta)}.
\end{equation}
The wavefield was then $8\times$ up-scaled using third-order spline interpolation and propagated a distance of $\Delta=\qty{0.5}{\meter}$ using a Fresnel propagator \cite{paganin2006} to create the propagated intensity $I_P(x,y) = |\Psi(x,y,z=\Delta)|^2$. To model an independent dark-field induced blurring, a simulated Fokker--Planck diffusion coefficient $D(x,y)$ at \qty{25}{\keV} was created by overlaying four projected spheres (\cref{fig:sim-simD}), and then scaled to \qty{30}{\keV} using an ad-hoc energy-dependence of $D \propto \lambda^3$ (see \cref{sec:determine_dep} for further justification and detail). The diffusion coefficients were converted to blur widths using $\sigma(x,y) = \sqrt{2D(x,y)}\Delta$ \cite{paganin2019, morgan2019}, and finally applied to the propagated intensities $I_P(x,y)$ using a local Gaussian-type diffusion:
\begin{equation}
    \begin{split}
        \label{eq:sim_DF}
    I(x,y) = \int_{-\infty}^\infty &\int_{-\infty}^\infty I_P(x',y') \\
    &\times g(x,y;x', y', \sigma(x', y')) \mathop{dx'} \mathop{dy'}, 
    \end{split}
\end{equation}
where $g(x,y;x',y',\sigma)$ is a 2D Gaussian probability density function centered at $(x',y')$, with standard deviation $\sigma$. 
The resulting simulated images are shown in \cref{fig:sim-I-E1,fig:sim-I-E2}. Where the dark-field signal is present, we see a slight reduction in contrast (see insets in \cref{fig:sim-I-E1,fig:sim-I-E2}). This effect is especially pronounced at the lower energy.    
As we base our reconstruction on the Fokker--Planck model, we deliberately decided not to also use a Fokker--Planck-based forward model for simulation. The approach of Fresnel propagation followed by local diffusion has low computational requirements, is straightforward to implement, and directly simulates the blurring effect seen in experimental propagation-based images. Alternatively, one could simulate dark-field by fully modelling the scattering microstructure. This could be done by ensemble-averaging over the propagated images from many random rapidly-varying complex refractive index maps, or by binning the recorded intensity from a high-resolution simulation that directly simulates microstructure.  

\begin{figure}
    \centering
    \includegraphics[width=\columnwidth]{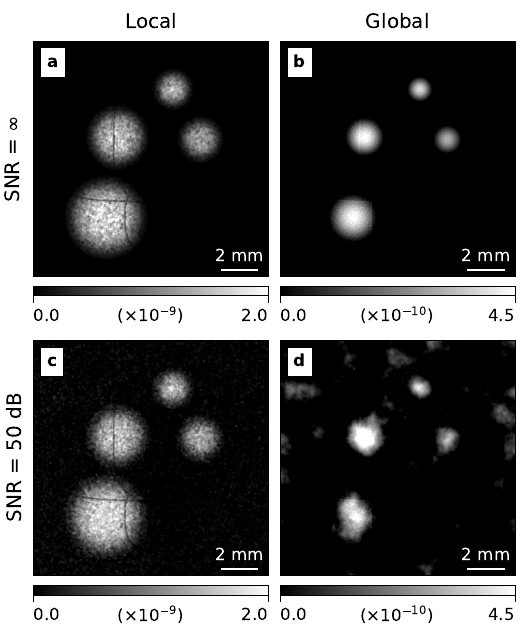}
    {\phantomsubcaption\label{fig:sim-D_local}}
    {\phantomsubcaption\label{fig:sim-D_global}}
    {\phantomsubcaption\label{fig:sim-D_local_noisy}}
    {\phantomsubcaption\label{fig:sim-D_global_noisy}}
    \caption{Reconstructions of the Fokker--Planck diffusion co-efficient (at \qty{25}{\keV}) from simulated data, using a global approach (\cref{eq:D_InvLap}) and a local approach (\cref{eq:D_from_vis}), with and without added noise. Note that the simulated diffusion coefficient had a range of \numrange{0}{2.5E-9} (\cref{fig:sim-simD}).}
    \label{fig:sim_res}
\end{figure}

\begin{figure}
    \centering
    \includegraphics[width=\columnwidth]{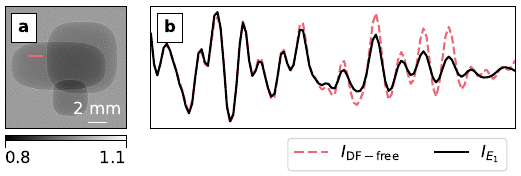}
    {\phantomsubcaption\label{fig:sim-Ivirt}}
    {\phantomsubcaption\label{fig:sim-IE1Ivirt}}
    \caption{(\subref{fig:sim-Ivirt}) $I_{\text{DF-free}}$ at \qty{25}{\keV}, created using the reconstructed simulated thickness in \cref{fig:sim-T}. (\subref{fig:sim-IE1Ivirt}) An intensity line profile taken along the red line in the first panel, which straddles strong and weak dark-field regions, shows that $I_{\text{DF-free}}$ recovers details that were lost to dark-field induced blurring in $I_{E_1}$ (\cref{fig:sim-I-E1}). The difference between these profiles is used in the global approach to reconstructing the diffusion coefficient (\cref{eq:D_InvLap}), and the change in the local visibility between these two profiles ($V_{\text{ref}}$ from $I_\text{DF-free}$ and $V_{\text{obs}}$ from $I_{E_1}$) gives a measure of dark-field via \cref{eq:D_from_vis} in the local approach.}
    \label{fig:sim_Ivirt}
\end{figure}

Using the resulting simulated intensity images, $I_{E_1}$(\cref{fig:sim-I-E1}) and $I_{E_2}$ (\cref{fig:sim-I-E2}), the projected thickness was reconstructed using \cref{eq:proj_thick,eq:iter_1}, with $n=20$ iterations. A convergence plot for this simulation, as well as for experimental data from \cref{sec:exp}, is shown in \cref{fig:iter}; both rapidly converge within approximately five iterations. The resulting projected thickness is shown in \cref{fig:sim-T}. In \cref{fig:sim-T_lp} we compare our result to TIE-based projected thickness reconstructions from the simulated images at both energies using an intensity line profile \cite{paganin2002}. Although the higher-energy image has the weaker dark-field effects, SPB-DF still achieves better fidelity than either of the TIE-based reconstructions by incorporating dark-field effects in the forward model. 

Based on the TIE propagator seen in \cref{eq:D_InvLap}, the dark-field-free image $I_{\text{DF-free}}$ at the lower energy $E_1=\qty{25}{\keV}$ was created using the reconstructed projected thickness. \Cref{fig:sim_Ivirt} compares the two key terms in \cref{eq:D_InvLap}, demonstrating a drop in local image contrast/visibility in the presence of dark-field, as seen in grating/speckle-based methods (e.g. \cite{pfeiffer2008, wen2010, berujon2012a}). Reconstructions of the diffusion coefficient $D(x,y)$ are shown in top row of \cref{fig:sim_res}, comparing the local (\cref{fig:sim-D_local}) and global (\cref{fig:sim-D_global}) approaches described in \cref{sec:theory}. A regularisation parameter of $\varepsilon = \num{2E5}$ was used in the global reconstruction. For the local reconstruction, visibility was measured in $10 \times 10$ pixel sliding windows using \cref{eq:vis}, and converted to the diffusion coefficient using \cref{eq:D_from_vis} with a period $p = 10~\text{pixels}$. The simulation was repeated with zero-mean Gaussian white noise added to the simulated images at a signal-to-noise ratio of $\text{SNR} = 10^{5}$, with the resulting reconstructions of the diffusion coefficient shown in the bottom row of \cref{fig:sim_res}. Note that all the reconstructions contain some values below zero, which are artefacts from the assumptions made in the theoretical derivation and from the naïve regularisation in global reconstruction. For our purposes `negative' diffusion (sharpening) is not considered physically relevant, and so we show all results with the lower grey value set to zero.       

Both methods give a reasonable qualitative reconstruction, and the local method is quantitatively close to the original signal. In all reconstructions there are some artefacts remaining from strong phase fringes, where the assumptions of the mathematical treatment in \cref{sec:theory} break down. As the local method relies on local windowing, the reconstruction looks noisy, even without added noise; by contrast, the global reconstruction is smooth. However, while the local reconstruction is quite robust to added noise, the global reconstruction is sensitive to low-frequency `cloud' artefacts that are typical for this kind of problem \cite{paganin2002}.

\section{Determining dark-field energy dependence}
\label{sec:determine_dep}

\begin{figure}
    \centering 
    \includegraphics[width=\columnwidth]{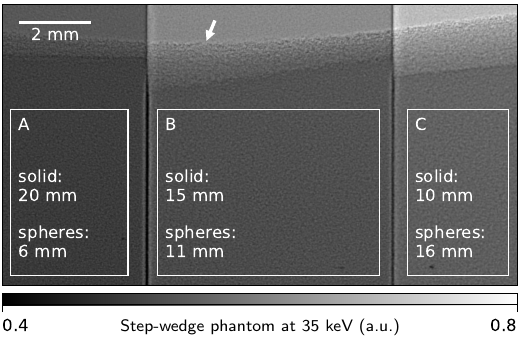}
    \caption{Regions-of-interest A--C shown on \qty{35}{\keV} PBI image of a PMMA step-wedge phantom. The regions consist of increasing depths of \qtyrange{45}{53}{\micro\meter} diameter PMMA microspheres (and decreasing depths of solid PMMA). All thicknesses have an uncertainty of \qty{\pm 1}{\mm}. The arrow marks the uneven top surface of the microspheres.}
    \label{fig:stepwedgeregions}
\end{figure}

\begin{figure}
    \centering
    \includegraphics[width=\columnwidth]{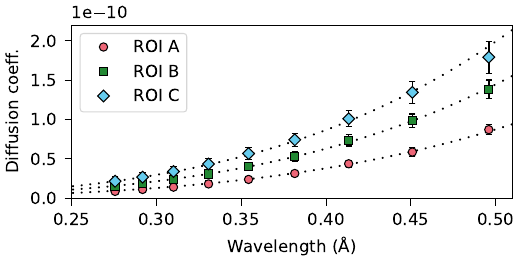}
    \caption{Mean Fokker--Planck diffusion coefficient $D$ in each region of interest (ROI) A--C of the step-wedge phantom (illustrated in \cref{fig:stepwedgeregions}) at each energy, measured using single-grid imaging. The error bars show the standard deviation in a region. The results are simultaneously fit to a power law model of the form $D_i(\lambda) = a_i\lambda^b$, where $\lambda$ is the wavelength, $i$ denotes each region, and $a_i$ and $b$ are four free parameters. The resulting energy dependence is $b = \num[separate-uncertainty = true]{3.72 \pm 0.03}$.}
    \label{fig:dfvsenergy}
\end{figure}

To apply our approach to experimental data, we must decide the Fokker--Planck diffusion coefficient's dependence on energy. To measure this dependence, a dark-field step-wedge phantom was imaged at several energies using single-grid imaging \cite{how2022} at the Imaging and Medical Beamline (IMBL) of the Australian Synchrotron in Melbourne, Australia. The beamline uses a superconducting multi-pole wiggler insertion device with \qty{1.4}{\tesla} field strength, and a bent double-Laue crystal monochromator was used to achieve an energy resolution of $\Delta E / E \sim 10^{-3}$ around the chosen beam energy \cite{stevenson2017}. The sample was placed approximately \qty{135}{\meter} downstream of the source, in imaging hutch 3B. An absorption grid, consisting of a stainless steel wire cloth with wire diameter of \qty{61}{\micro\meter} and hole size of \qty{90}{\micro\meter} (Test Sieve; ESSA), was placed upstream of the sample on a movable stage. The sample consisted of a poly(methyl~methacrylate) (PMMA) block, which had a central step-wedge shaped void cut out of it and was filled with PMMA microspheres with a diameter of \qtyrange{45}{53}{\micro\meter}  (DNP-P010; CD~Bioparticles). 
A sample-only image with annotated regions-of-interest (ROIs) is shown in \cref{fig:stepwedgeregions}. A \qty{2}{\meter} sample to detector distance was set. The detector used was IMBL's `Ruby', consisting of a \qty{25}{\micro\meter} Gd\textsubscript{2}O\textsubscript{2}S:Tb scintillator coupled to a PCO.edge sCMOS sensor (16-bit, $2560 \times 2160$~pixels) via a lens system, giving an effective pixel size of \qty{5.6}{\micro\meter}. 
The sample was imaged at energies of \qtylist[list-units=single]{25;27.5;30;32.5;35;37.5;40;42.5;45}{\keV}.

At each beam energy we recorded thirty grid-only, sample and grid, flat-field, and dark-current images. Images were averaged, and flat-field and dark-current corrected. We found that the grid and sample both independently moved up to five to ten pixels between energies. This could have been due to backlash error from repeated movement of the sample and grid stages into and out of the beam, as well as from the beam angle changing slightly when changing energy. Images were registered to the grid pattern using enhanced correlation coefficient (EEC) maximization \cite{evangelidis2008}. A small remaining movement of the sample between energies remained, but did not significantly affect the results as the final step in processing was an averaging over large regions-of-interest.

A single-grid reconstruction method was used to measure the dark-field signal in different ROIs of the sample at different energies \cite{how2022}. The method applies a local cross-correlation at each pixel between a small kernel in the sample-and-grid image, and a larger search region in the grid-only image. A sinusoid model of the grid is used to fit the data to the cross-correlation, giving the attenuation and relative change in visibility $\delta V$ (denoted `$DF$' in eq.~7 of \cite{how2022}) induced by the sample at each pixel. Finally, a median filter with size of the grid pitch is applied. The resulting reduction in $V$ was converted to a Fokker--Planck diffusion coefficient using \cref{eq:D_from_vis}. Note that this is equivalent to an initial conversion to a scattering angle $\theta$ according to eq.~11 in \cite{how2022}, which is directly related to the diffusion coefficient \cite{paganin2023}. 
The mean Fokker--Planck diffusion coefficient in each ROI measured this way is plotted against wavelength in \cref{fig:dfvsenergy}. To model the energy dependence of the diffusion coefficient a power law of the form $D(\lambda) = a\lambda^b$ was chosen, which has been demonstrated in theory \cite{sellerer2021,paganin2023}, simulation \cite{taphorn2021}, and experiment \cite{sellerer2021,taphorn2023} to be a good fit for dark-field diffusion energy-dependence in a variety of models (such as hard-spheres \cite{sellerer2021,paganin2023} and random walk through random media \cite{paganin2023}) and samples (both packed spheres and more complex microstructure \cite{sellerer2021,taphorn2021}). 
The model, with the four free parameters $a_i$ and $b$ (where $i\in \{1,2,3\}$ indexes the three ROIs), was fit to all measurements simultaneously using the Levenberg--Marquardt algorithm. The factors $a_i$ are expected to contain information about the projected thickness of microspheres in each region (alternatively, the number of scattering interfaces), and are assumed to be independent of energy. 
The resulting energy dependence was $b = \num[separate-uncertainty = true]{3.72 \pm 0.03}$. This result is broadly consistent with previous measurements of dark-field energy dependence in grating interferometry \cite{sellerer2021, taphorn2023}. An in-depth comparison would need to consider the difference in origin and measurement of dark-field signal, in both the propagation-based and interferometric context, but is beyond the scope of his article. 
This is an active area of research in the literature (see e.g. \cite{paganin2023}), and we emphasize that this dependence is likely to depend on details of the sample and experimental set-up. An updated understanding of dark-field energy dependence could be incorporated into $\Gamma(\lambda)$ without alteration to the body of the SPB-DF algorithm.    

\section{Experimental demonstration}
\label{sec:exp}

\begin{figure}
    \centering
    \includegraphics[width=\columnwidth]{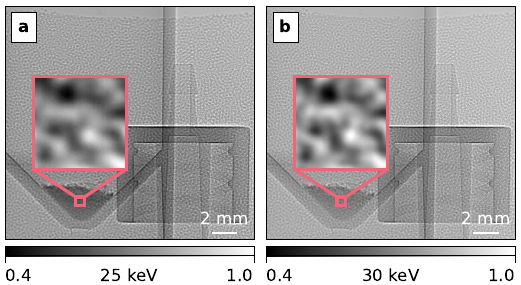}
    {\phantomsubcaption\label{fig:marc-E1}}
    {\phantomsubcaption\label{fig:marc-E2}}
    \caption{Experimental images of an approximately single-material sample at (\subref{fig:marc-E1}) \qty{25}{\keV} and (\subref{fig:marc-E2}) \qty{30}{\keV}. Most of the sample is composed of solid plastic, except for a powder of \qty{1}{\um} diameter polystyrene microspheres in the test tube at the bottom left of the images. There is a subtle but perceptible difference in the blurriness of the sample's texture in the region of the powder at the two energies.}
    \label{fig:marc-input}
\end{figure}

Experimental images were taken at IMBL, with the wiggler field strength adjusted to \qty{3}{\tesla}. The sample consisted of a plastic test tube containing polystyrene microspheres with a diameter of \qty{1}{\um} (P/N 100211-10; Corpuscular), placed next to a solid plastic male Luer lock adaptor. Directly in front of these was placed a custom made PMMA container, which was filled with \qtyrange{250}{300}{\um} diameter PMMA microspheres (Cospheric) to provide a consistent dominant local length scale. The container had walls of \qty{2}{\mm} thickness, with a \qty{1}{\mm} gap between them.       

As in \cref{sec:determine_dep}, the IMBL's `Ruby' detector was used, with an effective pixel size measured at \qty{9.7}{\um}. The sample was placed at a propagation distance of \qty{3.5}{\m}, and imaged at \qty{25}{\keV} and \qty{30}{\keV}. Images were flat-field and dark-current corrected, and registered using EEC maximization \cite{evangelidis2008}. The resulting images are shown in \cref{fig:marc-input}. As in the simulation, a slight blurring of the quickly-varying structure can be seen through the \qty{1}{\um} microspheres, which varies with energy. In a region containing only the large microspheres, a measurement of local length scale gave $p = 15.7$~pixels.     

\begin{figure}
    \centering
    \includegraphics[width=\columnwidth]{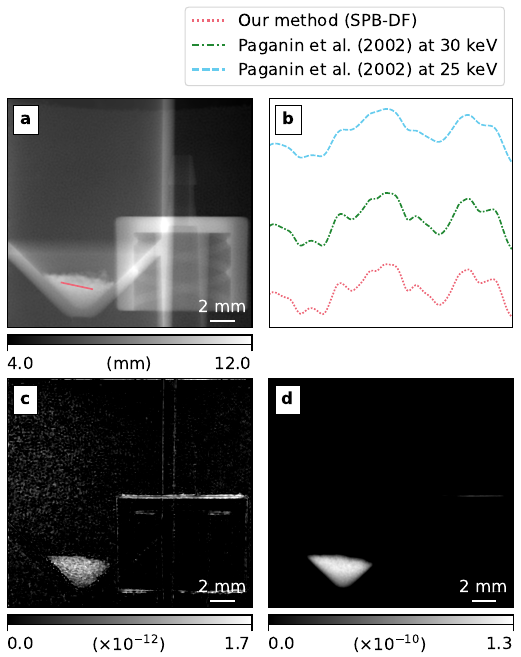}
    {\phantomsubcaption\label{fig:marc-T}}
    {\phantomsubcaption\label{fig:marc-T_lp}}
    {\phantomsubcaption\label{fig:marc-D_loc}}
    {\phantomsubcaption\label{fig:marc-D_glob}}
    \caption{Reconstructions from the experimental data. (\subref{fig:marc-T}) While all objects can be seen in the projected thickness reconstruction, only the \qty{1}{\um} diameter polystyrene microspheres are prominent in (\subref{fig:marc-D_loc}) local and (\subref{fig:marc-D_glob}) global dark-field reconstructions. While the local dark-field reconstruction contains some artefacts from the strong phase edges of the Luer lock, the global reconstruction is almost free of artefacts. (\subref{fig:marc-T_lp}) The projected thickness line profile along the red line shown in (\subref{fig:marc-T}) compared to projected thickness reconstructions using the TIE-based Paganin algorithm \cite{paganin2002} at each energy.}
    \label{fig:marc-output}
\end{figure}

The results of reconstructions of the experimental sample can be seen in \cref{fig:marc-output}. The dark-field energy dependence of $D \propto \lambda^{3.72}$ found in \cref{sec:determine_dep} was used in the reconstruction of the projected thickness (\cref{fig:marc-T}). As in the simulation, the thickness was also reconstructed using the TIE-based Paganin algorithm at each of the two energies \cite{paganin2002}. A line profile of $T$ through the region of the sample with the strongest dark-field signal is shown in \cref{fig:marc-T_lp}, comparing the thickness reconstructions. While, unlike in the simulation (\cref{fig:sim-T_lp}), the true sample thickness is unknown, the trend of an improved visibility of features in the projected thickness reconstruction is the same as in the simulation. For the global dark-field reconstruction (\cref{fig:marc-D_glob}), a regularisation parameter of $\varepsilon = \num{2.88E7}$ was found to be optimal. The global reconstruction clearly differentiates the strongly scattering \qty{1}{\um} microspheres from the rest of the sample, with almost no obvious artefacts. In contrast, the local reconstruction (\cref{fig:marc-D_loc}) retains some artefacts from the strong phase edges on the Luer lock and test tube. 

\section{Discussion}
\label{sec:discussion}

Our method of dark-field imaging involves two steps: (1) the projected thickness of the sample is reconstructed, taking into account dark-field effects; and (2) based on the projected thickness, the Fokker--Planck diffusion coefficient is reconstructed. This first step provides the opportunity to improve the resolution and feasibility of the microscopy of samples with porous or granular regions \cite{leatham2024}.
For the second step, we have considered two approaches. A global approach is to numerically solve the diffusion part of the Fokker--Plank equation, a Poisson equation. While we have used a straightforward spectral solution with a manually optimised Tihkonov regularisation, more advanced regularisation schemes or numerical methods of solving the Poisson equation which may be more appropriate for this kind of diffusive dark-field reconstruction could be explored in future work. In addition to this global approach, we have suggested an alternative approach that is inspired by structured-illumination techniques, based on comparing the visibility of local features. In this case, we compare the visibility of a small region around each pixel in one image that contains strong dark-field effects (the raw, lower-energy image) and another that contains none ($I_\text{DF-free}$, created using the reconstructed projected thickness). 

We must emphasise that a reconstruction based on local visibility relies on local structure being present in the images, and on this local structure varying over a few pixels. Quantitative local reconstruction depends on knowing the period of local intensity variation. In the simulation and experiment we deliberately included a rapidly-varying and consistent texture in the sample. However, a generic sample will not in general create a well-defined intensity variation of consistent size/period. This could be addressed by first estimating a local period (e.g. using a 2D wavelet transform), and then adjusting the window size and period at each pixel accordingly. In sec.~1 of the supplemental materials we repeat both the simulation and experiment carried out in this paper, but do not deliberately include quickly-varying features in the samples. While a global inversion of the Fokker--Planck equation is able to reconstruct dark-field even when there are no visibly blurred features, local reconstruction of the simulation breaks down completely, and is less robust in the experimental case.

In sec.~2 of the supplemental materials we perform an experiment with a sample composed of a variety of materials, which deviates strongly from the single-material assumption. While the global dark-field reconstruction has some difficulties in this case, the local reconstruction is still able to give a good qualitative picture of the strongly scattering components of the sample. In addition, based on the simulation carried out in \cref{sec:sim}, we saw that a local reconstruction can be more robust to noise than a global solution.  

Interest in the spectral behaviour of dark-field imaging has been spurred recently by progress in the development and proliferation of photon-counting energy-resolving X-ray detectors. A dichromatic or polychromatic source together with an energy-resolving detector would enable SPB-DF to be applied with a single exposure, significantly ameliorating issues with registration. A third energy bin raises the possibility of avoiding the single-material assumption, enabling separate reconstruction of the attenuation and refraction channels in multi-material samples. Multi-energy propagation-based imaging can be used to perform phase-retrieved material segmentation using a linearised transport-of-intensity equation \cite{schaff2020}. This could be used to mitigate the attenuation artefacts that remain in our dark-field retrieval for multi-material samples, or could be extended to include dark-field effects based on our work. Imaging at additional energies may also allow for the extension of SPB-DF to directional dark-field imaging \cite{jensen2010, kagias2016, smith2022, croughan2023} by making an appropriate substitution of the scalar diffusion coefficient $D$, such as by a symmetric rank-2 diffusion tensor \cite{pavlov2021, paganin2023} of the form 
\begin{equation}
    D \rightarrow
\begin{bmatrix}
D_{xx} & \frac{1}{2}D_{xy}\\
\frac{1}{2}D_{xy} & D_{yy}
\end{bmatrix}.
\end{equation}
The imaging regime (energy, sample size) described within this paper is currently used in small-animal biomedical research studies \cite{morgan2020}, suggesting this approach could be of benefit in that work. Dual-energy X-ray imaging systems are already in widespread clinical and security use, raising the possibility that SPB-DF could  be adopted for dark-field imaging in these contexts. To observe propagation-based dark-field effects, high local contrast would be helpful, which becomes more difficult at the higher energies and large pixel sizes used for human imaging. With a high-coherence micro-focus source and high-resolution detector, local image contrast can be achieved by imaging phase contrast speckle \cite{kitchen2004}. With a low-coherence source, one possibility is to overlay a high-contrast reference pattern, in the style of single-grid/speckle-based imaging, but without the explicit need for a mask-only reference image and the resulting strict set-up stability requirements. The resulting reconstruction would contain both the sample and the mask pattern, but this mask could then be subtracted out before image assessment. Commercial application would also require developing specific imaging protocols, including a choice of the dark-field energy dependence parameter. Further research is needed to assess if this is viable. 

\section{Conclusion}

We have seen that dark-field effects are observable in propagation-based X-ray imaging, and that their strength changes with energy. We have developed a dual-energy imaging method of reconstructing the projected thickness of a sample that accounts for these dark-field effects using the X-ray Fokker--Planck model, improving the accuracy of reconstruction compared to TIE-based thickness reconstruction. Following the thickness reconstruction, we considered two methods of recovering the Fokker--Planck diffusion coefficient to give a dark-field image. A global method using Fourier transforms gives smooth qualitative solutions with low artefacts, but is susceptible to noise. An alternative method, comparing the visibility of local sample features in a raw image and a generated dark-field-free image, produces stronger artefacts, but gives quantitative results and is robust to both noise and multi-material samples. Further research is needed to develop robust methods for the general inverse problem of diffusion retrieval in dark-field imaging. Multi-energy reconstruction opens the door to using energy-discriminating detectors for single-exposure and time-resolved dark-field PBI. 

\subsubsection*{Funding}
National Health and Medical Research Council (IMPACT); Australian Nuclear Science and Technology Organisation (18642); Australian Research Council (FT18010037, DP230101327).

\subsubsection*{Acknowledgments}
The work of Jannis Ahlers is supported by an Australian Government Research Training Program (RTP) Scholarship.  We thank David Paganin and Henriette Bast for useful discussions, and Ying Ying How for sharing her single-grid dark-field retrieval code \cite{how2022}. The experiments were completed at the Australian Synchrotron, part of ANSTO, under proposals 18515, 18648, and 21352. The authors are grateful for the help provided by Chris Hall, Daniel Hausermann, Anton Maksimenko, and Matthew Cameron, the beamline scientists at the Imaging and Medical Beamline.

\subsubsection*{Disclosures}
\noindent The authors declare no conflicts of interest.

\subsubsection*{Data availability} Data underlying the results presented in this paper are not publicly available at this time but may be obtained from the authors upon reasonable request.

\bibliographystyle{abbrv}
\bibliography{references}

\begin{thebibliography}{10}

\bibitem{alkadhi2022}
H.~Alkadhi, A.~Euler, D.~Maintz, and D.~Sahani, editors.
\newblock {\em Spectral {{Imaging}}: {{Dual-Energy}}, {{Multi-Energy}} and
  {{Photon-Counting CT}}}.
\newblock Medical {{Radiology}}. {Springer International Publishing}, {Cham},
  2022.

\bibitem{alloo2022}
S.~J. Alloo, D.~M. Paganin, K.~S. Morgan, M.~J. Kitchen, A.~W. Stevenson, S.~C.
  Mayo, H.~T. Li, B.~M. Kennedy, A.~Maksimenko, J.~C. Bowden, and K.~M. Pavlov.
\newblock Dark-field tomography of an attenuating object using intrinsic x-ray
  speckle tracking.
\newblock {\em Journal of Medical Imaging}, 9(3):031502, Feb. 2022.

\bibitem{alvarez1976}
R.~E. Alvarez and A.~Macovski.
\newblock Energy-selective reconstructions in {{X-ray}} computerised
  tomography.
\newblock {\em Physics in Medicine \& Biology}, 21(5):733, Sept. 1976.

\bibitem{aminzadeh2022}
A.~Aminzadeh, B.~D. Arhatari, A.~Maksimenko, C.~J. Hall, D.~Hausermann, A.~G.
  Peele, J.~Fox, B.~Kumar, Z.~Prodanovic, M.~Dimmock, D.~Lockie, K.~M. Pavlov,
  Y.~I. Nesterets, D.~Thompson, S.~C. Mayo, D.~M. Paganin, S.~T. Taba,
  S.~Lewis, P.~C. Brennan, H.~M. Quiney, and T.~E. Gureyev.
\newblock Imaging {{Breast Microcalcifications Using Dark-Field Signal}} in
  {{Propagation-Based Phase-Contrast Tomography}}.
\newblock {\em IEEE Transactions on Medical Imaging}, pages 1--1, 2022.

\bibitem{berujon2012a}
S.~Berujon, H.~Wang, and K.~Sawhney.
\newblock X-ray multimodal imaging using a random-phase object.
\newblock {\em Physical Review A}, 86(6):063813, Dec. 2012.

\bibitem{berujon2012}
S.~B{\'e}rujon, E.~Ziegler, R.~Cerbino, and L.~Peverini.
\newblock Two-{{Dimensional X-Ray Beam Phase Sensing}}.
\newblock {\em Physical Review Letters}, 108(15):158102, Apr. 2012.

\bibitem{braig2020}
E.-M. Braig, N.~Roiser, M.~A. Kimm, M.~Busse, J.~Andrejewski, J.~Scholz,
  C.~Petrich, A.~Gustschin, A.~Sauter, J.~Bodden, F.~Meurer, R.~Korbel,
  F.~Pfeiffer, J.~Herzen, and D.~Pfeiffer.
\newblock X-ray {{Dark-Field Radiography}}: {{Potential}} for {{Visualization}}
  of {{Monosodium Urate Deposition}}.
\newblock {\em Investigative Radiology}, 55(8):494, Aug. 2020.

\bibitem{cloetens1996}
P.~Cloetens, R.~Barrett, J.~Baruchel, J.-P. Guigay, and M.~Schlenker.
\newblock Phase objects in synchrotron radiation hard x-ray imaging.
\newblock {\em Journal of Physics D: Applied Physics}, 29(1):133, Jan. 1996.

\bibitem{croughan2023}
M.~K. Croughan, Y.~Y. How, A.~Pennings, and K.~S. Morgan.
\newblock Directional dark-field retrieval with single-grid x-ray imaging.
\newblock {\em Optics Express}, 31(7):11578--11597, Mar. 2023.

\bibitem{david2002}
C.~David, B.~N{\"o}hammer, H.~H. Solak, and E.~Ziegler.
\newblock Differential x-ray phase contrast imaging using a shearing
  interferometer.
\newblock {\em Applied Physics Letters}, 81(17):3287--3289, Oct. 2002.

\bibitem{dreier2020}
E.~S. Dreier, A.~Bergamaschi, G.~K. Kallon, R.~Br{\"o}nnimann, U.~L. Olsen,
  A.~Olivo, and M.~Endrizzi.
\newblock Tracking based, high-resolution single-shot multimodal x-ray imaging
  in the laboratory enabled by the sub-pixel resolution capabilities of the
  {{M\"ONCH}} detector.
\newblock {\em Applied Physics Letters}, 117(26):264101, Dec. 2020.

\bibitem{endrizzi2014}
M.~Endrizzi, P.~C. Diemoz, T.~P. Millard, J.~Louise~Jones, R.~D. Speller, I.~K.
  Robinson, and A.~Olivo.
\newblock Hard {{X-ray}} dark-field imaging with incoherent sample
  illumination.
\newblock {\em Applied Physics Letters}, 104(2):024106, Jan. 2014.

\bibitem{evangelidis2008}
G.~D. Evangelidis and E.~Z. Psarakis.
\newblock Parametric {{Image Alignment Using Enhanced Correlation Coefficient
  Maximization}}.
\newblock {\em IEEE Transactions on Pattern Analysis and Machine Intelligence},
  30(10):1858--1865, Oct. 2008.

\bibitem{gureyev2001}
T.~E. Gureyev, S.~Mayo, S.~W. Wilkins, D.~Paganin, and A.~W. Stevenson.
\newblock Quantitative {{In-Line Phase-Contrast Imaging}} with {{Multienergy X
  Rays}}.
\newblock {\em Physical Review Letters}, 86(25):5827--5830, June 2001.

\bibitem{gureyev2020}
T.~E. Gureyev, D.~M. Paganin, B.~Arhatari, S.~T. Taba, S.~Lewis, P.~C. Brennan,
  and H.~M. Quiney.
\newblock Dark-field signal extraction in propagation-based phase-contrast
  imaging.
\newblock {\em Physics in Medicine \& Biology}, 65(21):215029, Nov. 2020.

\bibitem{gureyev1998a}
T.~E. Gureyev and S.~W. Wilkins.
\newblock On {{X-ray}} phase retrieval from polychromatic images.
\newblock {\em Optics Communications}, 147(4):229--232, Feb. 1998.

\bibitem{how2022}
Y.~Y. How and K.~S. Morgan.
\newblock Quantifying the x-ray dark-field signal in single-grid imaging.
\newblock {\em Optics Express}, 30(7):10899--10918, Mar. 2022.

\bibitem{jensen2010}
T.~H. Jensen, M.~Bech, O.~Bunk, T.~Donath, C.~David, R.~Feidenhans'l, and
  F.~Pfeiffer.
\newblock Directional x-ray dark-field imaging.
\newblock {\em Physics in Medicine \& Biology}, 55(12):3317, May 2010.

\bibitem{kagias2016}
M.~Kagias, Z.~Wang, P.~{Villanueva-Perez}, K.~Jefimovs, and M.~Stampanoni.
\newblock {{2D-Omnidirectional Hard-X-Ray Scattering Sensitivity}} in a
  {{Single Shot}}.
\newblock {\em Physical Review Letters}, 116(9):093902, Mar. 2016.

\bibitem{kitchen2017}
M.~J. Kitchen, G.~A. Buckley, T.~E. Gureyev, M.~J. Wallace, N.~{Andres-Thio},
  K.~Uesugi, N.~Yagi, and S.~B. Hooper.
\newblock {{CT}} dose reduction factors in the thousands using {{X-ray}} phase
  contrast.
\newblock {\em Scientific Reports}, 7(1):15953, Nov. 2017.

\bibitem{kitchen2020}
M.~J. Kitchen, G.~A. Buckley, L.~T. Kerr, K.~L. Lee, K.~Uesugi, N.~Yagi, and
  S.~B. Hooper.
\newblock Emphysema quantified: Mapping regional airway dimensions using {{2D}}
  phase contrast {{X-ray}} imaging.
\newblock {\em Biomedical Optics Express}, 11(8):4176--4190, Aug. 2020.

\bibitem{kitchen2004}
M.~J. Kitchen, D.~Paganin, R.~A. Lewis, N.~Yagi, K.~Uesugi, and S.~T. Mudie.
\newblock On the origin of speckle in x-ray phase contrast images of lung
  tissue.
\newblock {\em Physics in Medicine and Biology}, 49(18):4335--4348, Sept. 2004.

\bibitem{kitchen2010}
M.~J. Kitchen, D.~M. Paganin, K.~Uesugi, B.~J. Allison, R.~A. Lewis, S.~B.
  Hooper, and K.~M. Pavlov.
\newblock X-ray phase, absorption and scatter retrieval using two or more phase
  contrast images.
\newblock {\em Optics Express}, 18(19):19994--20012, Sept. 2010.

\bibitem{labriet2021}
H.~Labriet, S.~Berujon, and E.~Brun.
\newblock X-ray imaging device and associated imaging method, Jan. 2021.

\bibitem{leatham2023}
T.~A. Leatham, D.~M. Paganin, and K.~S. Morgan.
\newblock X-ray dark-field and phase retrieval without optics, via the
  {{Fokker}}\textendash{{Planck}} equation.
\newblock {\em IEEE Transactions on Medical Imaging}, pages 1--1, 2023.

\bibitem{leatham2024}
T.~A. Leatham, D.~M. Paganin, and K.~S. Morgan.
\newblock X-ray phase and dark-field computed tomography without optical
  elements.
\newblock {\em Optics Express}, 32(3):4588--4602, Jan. 2024.

\bibitem{meinel2014a}
F.~G. Meinel, F.~Schwab, A.~Yaroshenko, A.~Velroyen, M.~Bech, K.~Hellbach,
  J.~Fuchs, T.~Stiewe, A.~O. Yildirim, F.~Bamberg, M.~F. Reiser, F.~Pfeiffer,
  and K.~Nikolaou.
\newblock Lung tumors on multimodal radiographs derived from grating-based
  {{X-ray}} imaging \textendash{} {{A}} feasibility study.
\newblock {\em Physica Medica}, 30(3):352--357, May 2014.

\bibitem{michelson1927}
A.~A. Michelson.
\newblock {\em Studies in {{Optics}}}.
\newblock University of {{Chicago}} Science Series. {The University of Chicago
  Press}, {Chicago, Ill.}, 1927.

\bibitem{momose2003}
A.~Momose, S.~Kawamoto, I.~Koyama, Y.~Hamaishi, K.~Takai, and Y.~Suzuki.
\newblock Demonstration of {{X-Ray Talbot Interferometry}}.
\newblock {\em Japanese Journal of Applied Physics}, 42(7B):L866, July 2003.

\bibitem{morgan2019}
K.~S. Morgan and D.~M. Paganin.
\newblock Applying the {{Fokker}}\textendash{{Planck}} equation to
  grating-based x-ray phase and dark-field imaging.
\newblock {\em Scientific Reports}, 9(1):17465, Nov. 2019.

\bibitem{morgan2011}
K.~S. Morgan, D.~M. Paganin, and K.~K.~W. Siu.
\newblock Quantitative single-exposure x-ray phase contrast imaging using a
  single attenuation grid.
\newblock {\em Optics Express}, 19(20):19781--19789, Sept. 2011.

\bibitem{morgan2012}
K.~S. Morgan, D.~M. Paganin, and K.~K.~W. Siu.
\newblock X-ray phase imaging with a paper analyzer.
\newblock {\em Applied Physics Letters}, 100(12):124102, Mar. 2012.

\bibitem{morgan2020}
K.~S. Morgan, D.~Parsons, P.~Cmielewski, A.~McCarron, R.~Gradl, N.~Farrow,
  K.~Siu, A.~Takeuchi, Y.~Suzuki, K.~Uesugi, M.~Uesugi, N.~Yagi, C.~Hall,
  M.~Klein, A.~Maksimenko, A.~Stevenson, D.~Hausermann, M.~Dierolf,
  F.~Pfeiffer, and M.~Donnelley.
\newblock Methods for dynamic synchrotron {{X-ray}} respiratory imaging in live
  animals.
\newblock {\em Journal of Synchrotron Radiation}, 27(1):164--175, Jan. 2020.

\bibitem{morgan2010}
K.~S. Morgan, K.~K.~W. Siu, and D.~M. Paganin.
\newblock The projection approximation and edge contrast for x-ray
  propagation-based phase contrast imaging of a cylindrical edge.
\newblock {\em Optics Express}, 18(10):9865--9878, May 2010.

\bibitem{olivo2021}
A.~Olivo.
\newblock Edge-illumination x-ray phase-contrast imaging.
\newblock {\em Journal of Physics: Condensed Matter}, 33(36):363002, July 2021.

\bibitem{paganin2006}
D.~Paganin.
\newblock {\em Coherent {{X-Ray Optics}}}.
\newblock Oxford {{Series}} on {{Synchrotron Radiation}}. {Oxford University
  Press}, {Oxford, New York}, Jan. 2006.

\bibitem{paganin2002}
D.~Paganin, S.~C. Mayo, T.~E. Gureyev, P.~R. Miller, and S.~W. Wilkins.
\newblock Simultaneous phase and amplitude extraction from a single defocused
  image of a homogeneous object.
\newblock {\em Journal of Microscopy}, 206(1):33--40, 2002.

\bibitem{paganin2020}
D.~M. Paganin, V.~{Favre-Nicolin}, A.~Mirone, A.~Rack, J.~Villanova, M.~P.
  Olbinado, V.~Fernandez, J.~C. da~Silva, and D.~Pelliccia.
\newblock Boosting spatial resolution by incorporating periodic boundary
  conditions into single-distance hard-x-ray phase retrieval.
\newblock {\em Journal of Optics}, 22(11):115607, Oct. 2020.

\bibitem{paganin2018}
D.~M. Paganin, H.~Labriet, E.~Brun, and S.~Berujon.
\newblock Single-image geometric-flow x-ray speckle tracking.
\newblock {\em Physical Review A}, 98(5):053813, Nov. 2018.

\bibitem{paganin2019}
D.~M. Paganin and K.~S. Morgan.
\newblock X-ray {{Fokker}}\textendash{{Planck}} equation for paraxial imaging.
\newblock {\em Scientific Reports}, 9(1):17537, Nov. 2019.

\bibitem{paganin2023}
D.~M. Paganin, D.~Pelliccia, and K.~S. Morgan.
\newblock Paraxial diffusion-field retrieval.
\newblock {\em Physical Review A}, 108(1):013517, July 2023.

\bibitem{pagot2003}
E.~Pagot, P.~Cloetens, S.~Fiedler, A.~Bravin, P.~Coan, J.~Baruchel,
  J.~H{\"a}rtwig, and W.~Thomlinson.
\newblock A method to extract quantitative information in analyzer-based x-ray
  phase contrast imaging.
\newblock {\em Applied Physics Letters}, 82(20):3421--3423, May 2003.

\bibitem{partridge2022}
T.~Partridge, A.~Astolfo, S.~S. Shankar, F.~A. Vittoria, M.~Endrizzi,
  S.~Arridge, T.~{Riley-Smith}, I.~G. Haig, D.~Bate, and A.~Olivo.
\newblock Enhanced detection of threat materials by dark-field x-ray imaging
  combined with deep neural networks.
\newblock {\em Nature Communications}, 13(1):4651, Sept. 2022.

\bibitem{pavlov2020a}
K.~M. Pavlov, H.~T. Li, D.~M. Paganin, S.~Berujon, H.~{Roug{\'e}-Labriet}, and
  E.~Brun.
\newblock Single-{{Shot X-Ray Speckle-Based Imaging}} of a {{Single-Material
  Object}}.
\newblock {\em Physical Review Applied}, 13(5):054023, May 2020.

\bibitem{pavlov2020}
K.~M. Pavlov, D.~M. Paganin, H.~T. Li, S.~Berujon, H.~{Roug{\'e}-Labriet}, and
  E.~Brun.
\newblock X-ray multi-modal intrinsic-speckle-tracking.
\newblock {\em Journal of Optics}, 22(12):125604, Nov. 2020.

\bibitem{pavlov2021}
K.~M. Pavlov, D.~M. Paganin, K.~S. Morgan, H.~T. Li, S.~Berujon, L.~Qu{\'e}not,
  and E.~Brun.
\newblock Directional dark-field implicit x-ray speckle tracking using an
  anisotropic-diffusion {{Fokker-Planck}} equation.
\newblock {\em Physical Review A}, 104(5):053505, Nov. 2021.

\bibitem{pelzer2014}
G.~Pelzer, A.~Zang, G.~Anton, F.~Bayer, F.~Horn, M.~Kraus, J.~Rieger,
  A.~Ritter, J.~Wandner, T.~Weber, A.~Fauler, M.~Fiederle, W.~S. Wong,
  M.~Campbell, J.~Meiser, P.~Meyer, J.~Mohr, and T.~Michel.
\newblock Energy weighted x-ray dark-field imaging.
\newblock {\em Optics Express}, 22(20):24507--24515, Oct. 2014.

\bibitem{pfeiffer2008}
F.~Pfeiffer, M.~Bech, O.~Bunk, P.~Kraft, E.~F. Eikenberry, C.~Br{\"o}nnimann,
  C.~Gr{\"u}nzweig, and C.~David.
\newblock Hard-{{X-ray}} dark-field imaging using a grating interferometer.
\newblock {\em Nature Materials}, 7(2):134--137, Feb. 2008.

\bibitem{quenot2021}
L.~Qu{\'e}not, H.~{Roug{\'e}-Labriet}, S.~Bohic, S.~Berujon, and E.~Brun.
\newblock Implicit tracking approach for {{X-ray}} phase-contrast imaging with
  a random mask and a conventional system.
\newblock {\em Optica}, 8(11):1412--1415, Nov. 2021.

\bibitem{rigon2007}
L.~Rigon, F.~Arfelli, and R.-H. Menk.
\newblock Generalized diffraction enhanced imaging to retrieve absorption,
  refraction and scattering effects.
\newblock {\em Journal of Physics D: Applied Physics}, 40(10):3077, May 2007.

\bibitem{schaff2020}
F.~Schaff, K.~S. Morgan, J.~A. Pollock, L.~C.~P. Croton, S.~B. Hooper, and
  M.~J. Kitchen.
\newblock Material {{Decomposition Using Spectral Propagation-Based
  Phase-Contrast X-Ray Imaging}}.
\newblock {\em IEEE Transactions on Medical Imaging}, 39(12):3891--3899, Dec.
  2020.

\bibitem{schaff2022}
F.~Schaff, J.~A. Pollock, K.~S. Morgan, and M.~J. Kitchen.
\newblock Spectral propagation-based x-ray phase-contrast computed tomography.
\newblock {\em Journal of Medical Imaging}, 9(3):031506, Mar. 2022.

\bibitem{schoonjans2011}
T.~Schoonjans, A.~Brunetti, B.~Golosio, M.~{Sanchez del Rio}, V.~A. Sol{\'e},
  C.~Ferrero, and L.~Vincze.
\newblock The xraylib library for {{X-ray}}\textendash matter interactions.
  {{Recent}} developments.
\newblock {\em Spectrochimica Acta Part B: Atomic Spectroscopy},
  66(11):776--784, Nov. 2011.

\bibitem{sellerer2021}
T.~Sellerer, K.~Mechlem, R.~Tang, K.~A. Taphorn, F.~Pfeiffer, and J.~Herzen.
\newblock Dual-{{Energy X-Ray Dark-Field Material Decomposition}}.
\newblock {\em IEEE Transactions on Medical Imaging}, 40(3):974--985, Mar.
  2021.

\bibitem{smith2022}
R.~Smith, F.~D. Marco, L.~Broche, M.-C. Zdora, N.~W. Phillips, R.~Boardman, and
  P.~Thibault.
\newblock X-ray directional dark-field imaging using {{Unified Modulated
  Pattern Analysis}}.
\newblock {\em PLOS ONE}, 17(8):e0273315, Aug. 2022.

\bibitem{smith-bindman2012}
R.~{Smith-Bindman}, D.~L. Miglioretti, E.~Johnson, C.~Lee, H.~S. Feigelson,
  M.~Flynn, R.~T. Greenlee, R.~L. Kruger, M.~C. Hornbrook, D.~Roblin, L.~I.
  Solberg, N.~Vanneman, S.~Weinmann, and A.~E. Williams.
\newblock Use of {{Diagnostic Imaging Studies}} and {{Associated Radiation
  Exposure}} for {{Patients Enrolled}} in {{Large Integrated Health Care
  Systems}}, 1996-2010.
\newblock {\em JAMA}, 307(22):2400--2409, June 2012.

\bibitem{snigirev1995}
A.~Snigirev, I.~Snigireva, V.~Kohn, S.~Kuznetsov, and I.~Schelokov.
\newblock On the possibilities of x-ray phase contrast microimaging by coherent
  high-energy synchrotron radiation.
\newblock {\em Review of Scientific Instruments}, 66(12):5486--5492, Dec. 1995.

\bibitem{speller1983}
R.~D. Speller, G.~J. Ensell, and C.~Wallis.
\newblock A system for dual-energy radiography.
\newblock {\em The British Journal of Radiology}, 56(667):461--465, July 1983.

\bibitem{stevenson2017}
A.~W. Stevenson, J.~C. Crosbie, C.~J. Hall, D.~H{\"a}usermann, J.~Livingstone,
  and J.~E. Lye.
\newblock Quantitative characterization of the {{X-ray}} beam at the
  {{Australian Synchrotron Imaging}} and {{Medical Beamline}} ({{IMBL}}).
\newblock {\em Journal of Synchrotron Radiation}, 24(1):110--141, Jan. 2017.

\bibitem{taphorn2020a}
K.~Taphorn, F.~De~Marco, J.~Andrejewski, T.~Sellerer, F.~Pfeiffer, and
  J.~Herzen.
\newblock Grating-based spectral {{X-ray}} dark-field imaging for correlation
  with structural size measures.
\newblock {\em Scientific Reports}, 10(1):13195, Aug. 2020.

\bibitem{taphorn2023}
K.~Taphorn, L.~Kaster, T.~Sellerer, A.~H{\"o}tger, and J.~Herzen.
\newblock Spectral {{X-ray}} dark-field signal characterization from
  dual-energy projection phase-stepping data with a {{Talbot-Lau}}
  interferometer.
\newblock {\em Scientific Reports}, 13(1):767, Jan. 2023.

\bibitem{taphorn2021}
K.~Taphorn, K.~Mechlem, T.~Sellerer, F.~De~Marco, M.~Viermetz, F.~Pfeiffer,
  D.~Pfeiffer, and J.~Herzen.
\newblock Direct {{Differentiation}} of {{Pathological Changes}} in the {{Human
  Lung Parenchyma With Grating-Based Spectral X-ray Dark-Field Radiography}}.
\newblock {\em IEEE Transactions on Medical Imaging}, 40(6):1568--1578, June
  2021.

\bibitem{teague1983}
M.~R. Teague.
\newblock Deterministic phase retrieval: A {{Green}}'s function solution.
\newblock {\em Journal of the Optical Society of America}, 73(11):1434--1441,
  Nov. 1983.

\bibitem{wang2014}
Z.~Wang, N.~Hauser, G.~Singer, M.~Trippel, R.~A. {Kubik-Huch}, C.~W. Schneider,
  and M.~Stampanoni.
\newblock Non-invasive classification of microcalcifications with
  phase-contrast {{X-ray}} mammography.
\newblock {\em Nature Communications}, 5(1):3797, May 2014.

\bibitem{wen2010}
H.~H. Wen, E.~E. Bennett, R.~Kopace, A.~F. Stein, and V.~Pai.
\newblock Single-shot x-ray differential phase contrast and diffraction imaging
  using two-dimensional transmission gratings.
\newblock {\em Optics letters}, 35(12):1932--1934, June 2010.

\bibitem{wilkins1996}
S.~W. Wilkins, T.~E. Gureyev, D.~Gao, A.~Pogany, and A.~W. Stevenson.
\newblock Phase-contrast imaging using polychromatic hard {{X-rays}}.
\newblock {\em Nature}, 384(6607):335--338, Nov. 1996.

\bibitem{yang2014}
F.~Yang, F.~Prade, M.~Griffa, I.~Jerjen, C.~Di~Bella, J.~Herzen, A.~Sarapata,
  F.~Pfeiffer, and P.~Lura.
\newblock Dark-field {{X-ray}} imaging of unsaturated water transport in porous
  materials.
\newblock {\em Applied Physics Letters}, 105(15):154105, Oct. 2014.

\bibitem{yaroshenko2015}
A.~Yaroshenko, K.~Hellbach, A.~O. Yildirim, T.~M. Conlon, I.~E. Fernandez,
  M.~Bech, A.~Velroyen, F.~G. Meinel, S.~Auweter, M.~Reiser, O.~Eickelberg, and
  F.~Pfeiffer.
\newblock Improved {{In}} vivo {{Assessment}} of {{Pulmonary Fibrosis}} in
  {{Mice}} using {{X-Ray Dark-Field Radiography}}.
\newblock {\em Scientific Reports}, 5(1):17492, Dec. 2015.

\bibitem{zdora2018}
M.-C. Zdora.
\newblock State of the {{Art}} of {{X-ray Speckle-Based Phase-Contrast}} and
  {{Dark-Field Imaging}}.
\newblock {\em Journal of Imaging}, 4(5):60, May 2018.

\bibitem{zdora2017}
M.-C. Zdora, P.~Thibault, T.~Zhou, F.~J. Koch, J.~Romell, S.~Sala, A.~Last,
  C.~Rau, and I.~Zanette.
\newblock X-ray {{Phase-Contrast Imaging}} and {{Metrology}} through {{Unified
  Modulated Pattern Analysis}}.
\newblock {\em Physical Review Letters}, 118(20):203903, May 2017.

\bibitem{zhong2000}
Z.~Zhong, W.~Thomlinson, D.~Chapman, and D.~Sayers.
\newblock Implementation of diffraction-enhanced imaging experiments: At the
  {{NSLS}} and {{APS}}.
\newblock {\em Nuclear Instruments and Methods in Physics Research Section A:
  Accelerators, Spectrometers, Detectors and Associated Equipment},
  450(2):556--567, Aug. 2000.

\end{thebibliography}


\begin{thebibliography}{1}

\bibitem{evangelidis2008}
Georgios~D. Evangelidis and Emmanouil~Z. Psarakis.
\newblock Parametric {{Image Alignment Using Enhanced Correlation Coefficient
  Maximization}}.
\newblock {\em IEEE Transactions on Pattern Analysis and Machine Intelligence},
  30(10):1858--1865, October 2008.

\bibitem{morgan2019}
Kaye~S. Morgan and David~M. Paganin.
\newblock Applying the {{Fokker}}\textendash{{Planck}} equation to
  grating-based x-ray phase and dark-field imaging.
\newblock {\em Scientific Reports}, 9(1):17465, November 2019.

\bibitem{paganin2002}
D.~Paganin, S.~C. Mayo, T.~E. Gureyev, P.~R. Miller, and S.~W. Wilkins.
\newblock Simultaneous phase and amplitude extraction from a single defocused
  image of a homogeneous object.
\newblock {\em Journal of Microscopy}, 206(1):33--40, 2002.

\bibitem{preibisch2009}
Stephan Preibisch, Stephan Saalfeld, and Pavel Tomancak.
\newblock Globally optimal stitching of tiled {{3D}} microscopic image
  acquisitions.
\newblock {\em Bioinformatics}, 25(11):1463--1465, June 2009.

\bibitem{schoonjans2011}
Tom Schoonjans, Antonio Brunetti, Bruno Golosio, Manuel {Sanchez del Rio},
  Vicente~Armando Sol{\'e}, Claudio Ferrero, and Laszlo Vincze.
\newblock The xraylib library for {{X-ray}}\textendash matter interactions.
  {{Recent}} developments.
\newblock {\em Spectrochimica Acta Part B: Atomic Spectroscopy},
  66(11):776--784, November 2011.

\end{thebibliography}

\end{document}


\maketitle

\section{Dark-field reconstruction without local contrast}

In this paper, we presented two methods of reconstructing the X-ray Fokker--Planck diffusion coefficient from our reconstruction of the projected sample thickness. The `local' approach, inspired by structured-illumination techniques, quantifies the dark-field signal using a difference in visibility of local features of the sample in the raw low-energy image and the dark-field-free image derived from the projected thickness. This local approach is reliant on local structure existing in the images. As diffusive dark-field is primarily associated with random unresolved micro-structure, such local structure is often present in coherent X-ray images in the form of interferometric speckle. Nevertheless, it is interesting to consider the possibility of weak local contrast. This may be the case for large pixel sizes, or when such contrast is hidden by noise. It is tempting to think that dark-field effects would then not be present, as the visibility of non-visible structure obviously cannot be reduced. However, this may not give a full picture. Consider the expansion of the diffusion term in the X-ray Fokker--Planck equation:
\begin{equation}
\label{eq:FP-diffterm}
    \nabla_\perp^2 (D I_0) = D \nabla_\perp^2 I_0 + I_0 \nabla_\perp^2 D + 2 \nabla_\perp D \cdot \nabla_\perp I_0, 
\end{equation}
where $I_0$ is the intensity at the exit-surface of the sample. Without local contrast, the first and last terms will be strongly suppressed. However, a changing dark-field can lead to a transport of intensity via the middle term, proportional to $\nabla_\perp^2 D(x,y)$ (see also fig.~4 and associated text in \cite{morgan2019}). 

\subsection{Simulation}

\begin{figure}
\centering{\includegraphics{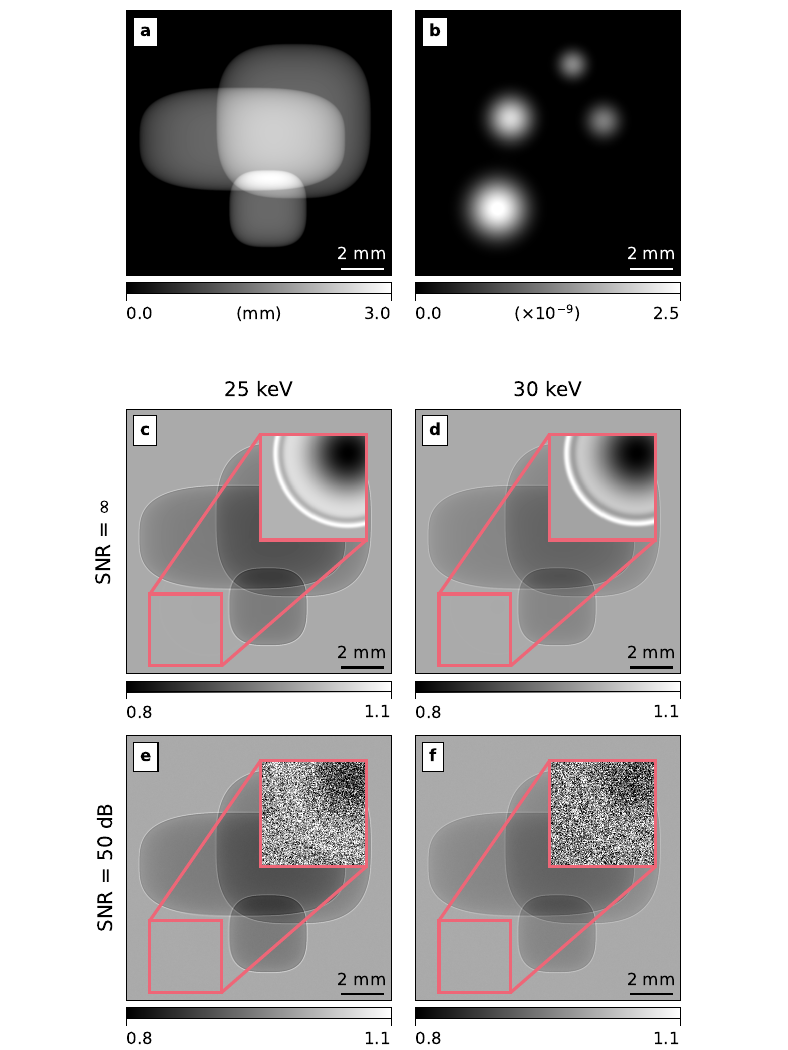}
  \phantomsubcaption\label{fig:sim-simT}
  \phantomsubcaption\label{fig:sim-simD}
  \phantomsubcaption\label{fig:sim-I-E1}
  \phantomsubcaption\label{fig:sim-I-E2}
  \phantomsubcaption\label{fig:sim-I-E1_noisy}
  \phantomsubcaption\label{fig:sim-I-E2_noisy}}
\captionsetup{singlelinecheck=off}
\caption[]{Simulated data without high-frequency features. (\subref{fig:sim-simT})~$T_\text{in}(x,y)$, (\subref{fig:sim-simD})~$D(x,y)$ (at \qty{25}{\keV}), and the propagated and blurred intensities at \qty{25}{\keV} and \qty{30}{\keV}, with and without added noise. As there are no quickly varying features, no blurring of such features is evident in the images. It is therefore difficult to immediately see any effects of dark-field diffusion. However, the effects can be seen by significantly adjusting the greyscale. The insets show a region of the image with a strong and changing dark-field signal. The greyscale range of these insets is the maximum and minimum within that region at each energy in the zero-noise images:\\
\qty{25}{\keV}: [0.99852, 1.00064]\\
\qty{30}{\keV}: [0.99914, 1.00048]
}
\label{fig:sim_sim-data}
\end{figure}

\begin{figure}
\centering{\includegraphics{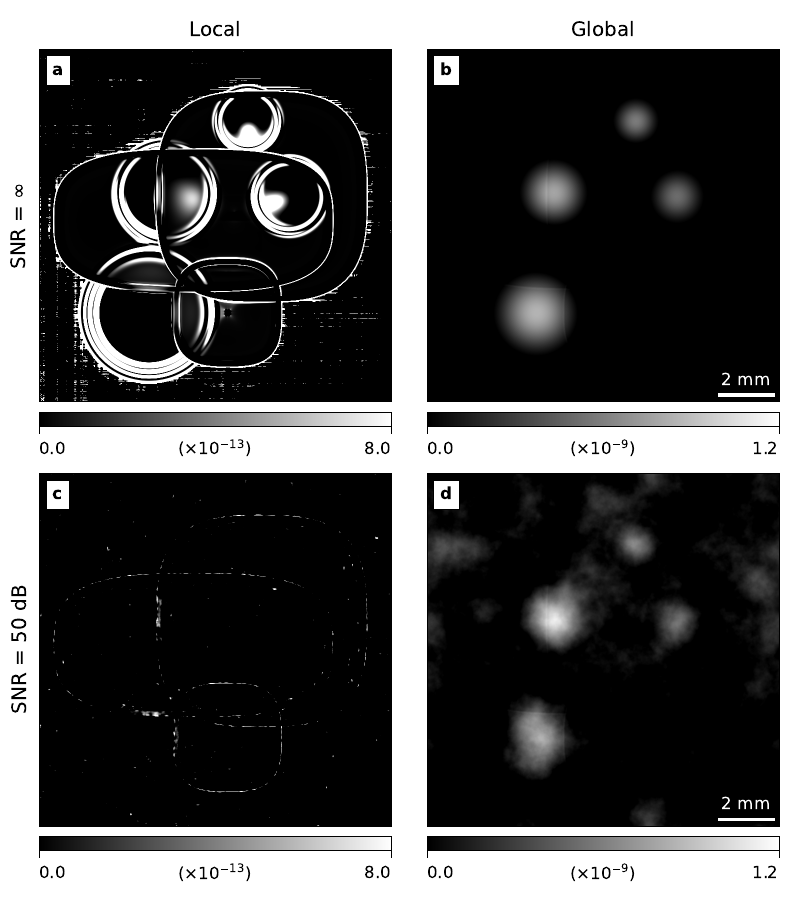}
  \phantomsubcaption\label{fig:sim-D_local}
  \phantomsubcaption\label{fig:sim-D_global}
  \phantomsubcaption\label{fig:sim-D_local_noisy}
  \phantomsubcaption\label{fig:sim-D_global_noisy}}
\caption{Reconstructions of $D(x,y)$ (at \qty{25}{\keV}) from simulated data without high-frequency features. Unsurprisingly, local reconstructions fail. (\subref{fig:sim-D_local}) Without any added noise some small remnants of the transport of intensity by the dark field can be seen, but (\subref{fig:sim-D_local_noisy}) these disappear when noise is added. In contrast, a global reconstruction gives good qualitative results, even when noise is added.}
\label{fig:sim_res}
\end{figure}

With this in mind, we repeated our simulation, but did not include the quickly-varying texture (from a random absorption mask) in the simulated projected thickness $T_\text{in}$. Note that one consequence of this was that some of the simulated dark-field signal was now in regions with zero simulated thickness, effectively simulating a `pure-dark-field' condition. Apart from changing the projected thickness, all other simulation parameters were kept the same. The simulated data is shown in \cref{fig:sim_sim-data}. Looking at the simulated images \cref{fig:sim-I-E1,fig:sim-I-E2}, it is not immediately apparent that dark-field effects are present. However, adjusting the greyscale in regions-of-interest around a strong dark-field signal reveals that there has been some small transport of intensity. As the small-angle scattering process (here simulated with a local Gaussian-type diffusion) pushes intensity away from regions of strong scattering, we observe an extinction of intensity near the centre of the large projected sphere in the simulated $D(x,y)$, and in turn an increase in intensity near the edge, giving a `glowing' appearance. As in sec.~3 of the main text, the simulation was then repeated with noise added at a signal-to-noise ratio of $\text{SNR} = \qty{50}{\dB}$). Some remnant of the effect of transport of intensity due to dark-field diffusion can still be seen in the inserts in the simulated images at (\cref{fig:sim-I-E1_noisy}) \qty{25}{\keV} and (\cref{fig:sim-I-E2_noisy}) \qty{30}{\keV}, but it is nearly swamped by noise.      

Global and local reconstructions of $D(x,y)$ are shown in \cref{fig:sim_res}. For the global reconstruction, a regularisation parameter of $\varepsilon = \num{1.3E6}$ was used; all other reconstruction parameters were kept the same. Unsurprisingly, local reconstruction breaks down, although some remnants of the diffusion are still evident when there is no noise (\cref{fig:sim-D_local}). On the other hand, a global solution to the full Fokker--Planck equation (\cref{fig:sim-D_global}) gives excellent qualitative results. Even when the transport of intensity due to the term proportional to $I_0 \nabla_\perp^2 D$ is nearly hidden by noise (as can be seen in \cref{fig:sim-I-E1_noisy,fig:sim-I-E2_noisy}), a global reconstruction can give a fair result (\cref{fig:sim-D_global_noisy}).         

\subsection{Experiment}

\begin{figure}
\centering{\includegraphics{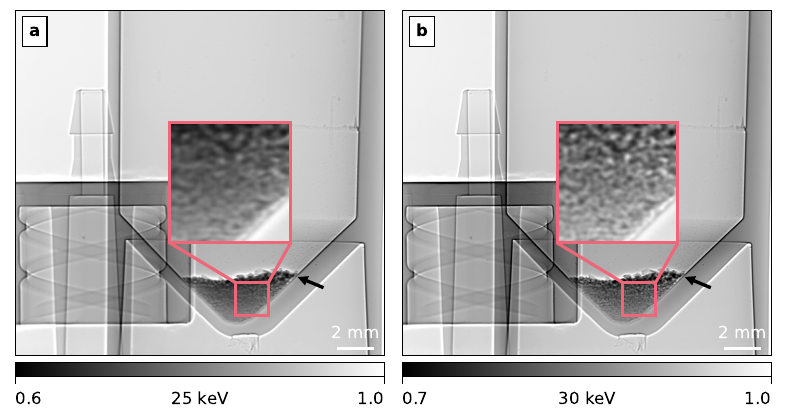}
  \phantomsubcaption\label{fig:jannis-E1}
  \phantomsubcaption\label{fig:jannis-E2}}
\caption{Propagation-based images at two energies of a near single-material sample that does not produce a consistent texture over the entire image. The sample consists of \qty{1}{\um} polystyrene microspheres in a plastic test tube, placed next to a solid plastic male Luer lock adaptor. While most of the sample is very smooth, the microspheres produce some texture. In addition, there appears to be a `glowing' effect around the microspheres, similar to the one seen in simulation (\cref{fig:sim-I-E1,fig:sim-I-E2}). The arrows indicate where the microspheres begin. Note the sharp transition of the strong, dark phase edge of the inside edge of the test tube (above the arrow), which seems to be overpowered by the glow around the microspheres below the arrow.}
\label{fig:jannis_input}
\end{figure}

\begin{figure}
\centering{\includegraphics{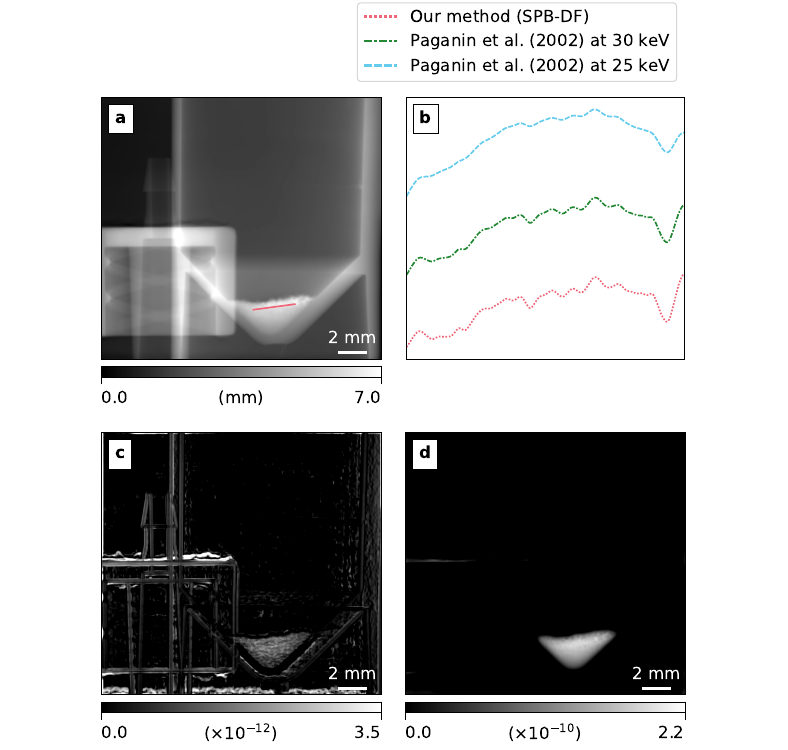}
  \phantomsubcaption\label{fig:jannis-T}
  \phantomsubcaption\label{fig:jannis-T_lp}
  \phantomsubcaption\label{fig:jannis-D_local}
  \phantomsubcaption\label{fig:jannis-D_global}}
\caption{Reconstruction results of a smooth experimental sample. (\subref{fig:jannis-T}) All objects are visible in the reconstructed projected thickness, while (\subref{fig:jannis-D_global}) only the highly scattering \qty{1}{\um} microspheres can be seen in the global dark-field reconstruction. (\subref{fig:jannis-D_local}) As the microspheres produce some local texture, they are visible in the local dark-field reconstruction, but the lack of a consistent texture leads to significant artefacts from solid features in the sample. (\subref{fig:jannis-T_lp}) A comparison of the line profile shown in (\subref{fig:jannis-T}) with projected thickness reconstruction using a TIE-based Paganin algorithm \cite{paganin2002} at both \qty{25}{\keV} and \qty{30}{\keV}.}
\label{fig:jannis_output}
\end{figure}

The experimental sample used in sec.~5 of the main text was deliberately chosen to create a `texture' with a consistent length scale. In a local reconstruction of the diffusion coefficient, this allowed the use of a single value for the period $p$ and window size throughout the whole image. To demonstrate our method on a smooth sample, we repeated the experiment, but this time did not include the container of large \qtyrange{250}{300}{\um} diameter PMMA microspheres. All other experimental parameters were kept the same. The resulting flat-field and dark-current corrected images can be seen in \cref{fig:jannis_input}. The images are overall much smoother, but we note that there is still some grainy texture within the \qty{1}{\um} polystyrene microspheres. Although the microspheres themselves are much smaller than the pixel size, some clumping could have occurred that lead to resolvable texture. In addition, we may be seeing some interferometric near-field speckle. 

Reconstructions were performed with the same parameters as in the main text. The results are shown in \cref{fig:jannis_output}. Even with no consistent local texture, the global reconstruction of $D$ (\cref{fig:jannis-D_global}) clearly differentiates the scattering microspheres from the rest of the sample. Because of the texture created by the microspheres, some dark-field signal is still seen in that region in the local reconstruction (\cref{fig:jannis-D_local}). However, artefacts from the rest of the sample are much stronger than seen from the local reconstruction in the main text, and the signal within the microspheres is unlikely to be quantitatively accurate.   

\section{Multi-material sample}

\begin{figure}
    \centering
    \includegraphics{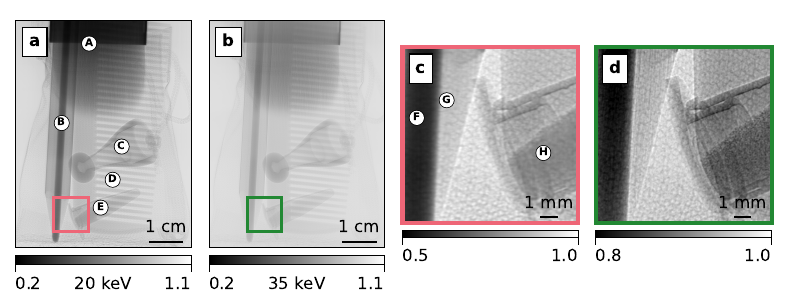}
    {\phantomsubcaption\label{fig:bag-I_E1}}
    {\phantomsubcaption\label{fig:bag-I_E2}}
    {\phantomsubcaption\label{fig:bag-I_E1_ROI}}
    {\phantomsubcaption\label{fig:bag-I_E2_ROI}}
    \caption{Propagation-based images of a mesh-cloth bag containing a make-up brush (A), pencil (B), playing piece (C), comb (D), and microcentrifuge tube containing \qty{6}{\micro\meter} PMMA microspheres (E). The sample was imaged at (\subref{fig:bag-I_E1}) \qty{20}{\keV} and (\subref{fig:bag-I_E2}) \qty{35}{\keV}. The zoomed section shows three strongly scattering parts of the sample: the graphite core (F) and wood fibres (G) in the pencil, and the PMMA microspheres (H). Compared to the higher energy image (\subref{fig:bag-I_E2_ROI}), the texture of the bag and the speckle pattern created by the microspheres have been strongly smoothed in these regions in the lower energy image (\subref{fig:bag-I_E1_ROI}). Note that the greyscales in (\subref{fig:bag-I_E1_ROI}) and (\subref{fig:bag-I_E2_ROI}) have been adjusted to better show the difference in visibility of the local structure.} 
    \label{fig:bag_input}
\end{figure}

\subsection{Method}

To assess how the dark-field imaging technique performs under the more challenging conditions of a sample with multiple different materials, a sample consisting of a variety of items was imaged at IMBL with a propagation distance of \qty{2}{\meter}. The sample consisted of a cloth bag, filled with an assortment of items, including: a pencil; a plastic playing piece; a make-up brush; a plastic hair comb; and a polypropylene microcentrifuge tube filled with \qty{6}{\micro\meter} diameter PMMA microspheres (Polybead 26305-500; Polysciences). The sample was imaged at \qty{20}{\keV} and \qty{35}{\keV}. The Ruby detector was used with an effective pixel size of \qty{20}{\um}. Due to the restricted size of the beam, the sample was imaged at two vertical positions. The images were registered using EEC maximization \cite{evangelidis2008}, and stitched with a linear blending \cite{preibisch2009}.

Experimental parameters were chosen by adjusting the propagation distance at the lower energy $E_1$ until some weak but apparent dark-field-induced blurring was observed, and then raising the energy until this blurring was significantly reduced, giving the second energy $E_2$. The registered and stitched images are shown in \cref{fig:bag-I_E1,fig:bag-I_E2}, with visible energy-related differences in the local visibility seen in the magnified regions shown in \cref{fig:bag-I_E1_ROI,fig:bag-I_E2_ROI}, particularly behind the pencil and microspheres. For the reconstruction, the energy dependence of $D \propto \lambda^{3.72}$ measured in section~4 of the main text was used. As many of the objects in the sample were composed of similar plastics, the refractive index of PMMA at the imaging energies (sourced from \texttt{xraylib} \cite{schoonjans2011}) was used in the algorithm. The SPB-DF thickness retrieval algorithm was iterated twenty times, and rapidly converged.

\subsection{Results}

\begin{figure}
    \centering
    \includegraphics{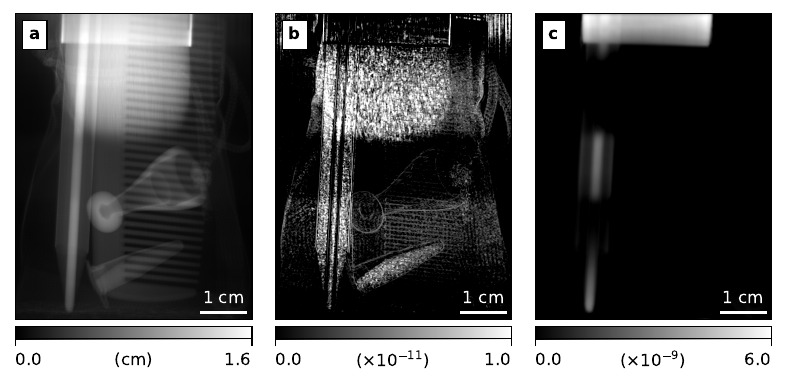}
    {\phantomsubcaption\label{fig:bag-T}}
    {\phantomsubcaption\label{fig:bag-D_local}}
    {\phantomsubcaption\label{fig:bag-D_global}}
    \caption{Reconstruction results from imaging a cloth bag containing sundry items. In (\subref{fig:bag-T}) the projected thickness, all items can be seen, while in (\subref{fig:bag-D_local}) the local dark-field reconstruction only the strongly scattering items are pronounced. (\subref{fig:bag-D_global}) A global dark-field reconstruction fails due to strong attenuation by both the metal ferrule of the make-up brush and the dense graphite powder in the pencil.}    
    \label{fig:bag_output}
\end{figure}

The recovered projected thickness and diffusion coefficients are shown in \cref{fig:bag_output}.  The threads of the bag created a fairly regular intensity pattern (clearly visible in \cref{fig:bag-I_E2_ROI}), and an auto-correlation of a region of the image containing only the bag was used to measure a period of $p~\approx~14~\text{pixels}$. As this sample is far from a weakly attenuating, single-material object, inversion of the Fokker--Planck equation is challenging. Notable exceptions to the single-material assumption were the wood and graphite of the pencil, and the metallic ferrule of the make-up brush. Even with an optimal regularisation of $\varepsilon = \num{1.7E7}$, these objects dominate the global reconstruction (\cref{fig:bag-D_global}). By contrast, the local dark-field reconstruction gives reasonable quantitative results, and clearly differentiates the highly scattering microsphere-filled microcentrifuge tube, hairs of the make-up brush, and wood grain and compressed graphite powder in the pencil (\cref{fig:bag-D_local}). Compared to the projected thickness (\cref{fig:bag-T}), the solid plastic playing piece and comb are suppressed. Some artefacts are still seen, particularly at strong phase edges. Note that some of the artefacts may be related to harmonic contamination, which is explained in \cref{sec:banding}. Overall, local reconstruction of the diffusion coefficient appears remarkably resilient to noise and deviation from the assumptions of a single-material sample.

\subsection{Banding artefacts}
\label{sec:banding}

\begin{figure}
 \centering
 {\includegraphics{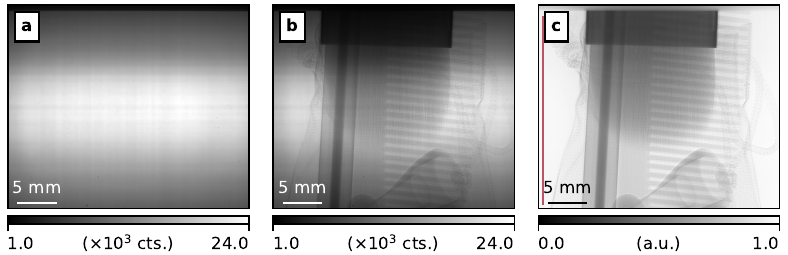}
  \phantomsubcaption\label{fig:art-flat}
  \phantomsubcaption\label{fig:art-im}
  \phantomsubcaption\label{fig:art-fad}}
\caption{Flat-field processing of \qty{20}{\keV} image of the upper half of the sample. (\subref{fig:art-flat}) The flat field shows strong central banding, potentially aggravated by third-harmonic contamination from the bent-Laue monochromator. (\subref{fig:art-im}) When the sample is introduced, the flat-field is distorted by a large detector point-spread function (PSF), leading to a banding artefact remaining in (\subref{fig:art-fad}) flat-field-corrected images (see \cref{fig:art-lp} for the line profile in (\subref{fig:art-fad})).}
\label{fig:artefacts}
\end{figure}

\begin{figure}
 \centering{\includegraphics{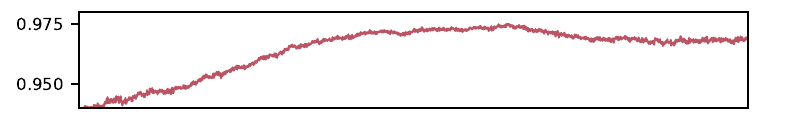}}
\caption{Intensity line profile through air in the flat-field and dark-current corrected image of the upper half of sample (\cref{fig:art-fad}). Note the variation of about \qty{4}{\percent}, which is amplified in processing and produces the streak artefacts observed in the reconstructed dark field.}
\label{fig:art-lp}
\end{figure}

In the dark-field reconstructions of the multi-material sample there are artefacts that resemble horizontal bands. We would like to emphasise that these artefacts are not inherently a product of the reconstruction, and are present in the raw data. The flat-field correction process of a raw recorded image at \SI{20}{\keV} of the upper half of the sample is shown in \cref{fig:artefacts}. The flat-field contains a strong central intensity band. We theorise that harmonic contamination in the monochromator led to a central band that included a non-trivial contribution from the third harmonic energy (60keV). In addition, a large detector point-spread function led to a remnant banding artefact in the flat-field corrected image, which is directly evident in a vertical line profile taken through a region of air next to the sample (\cref{fig:art-lp}). Note that the images of the sample in the paper were created by stitching two images of the top and bottom half of the sample vertically. Therefore, the single central band artefact in \cref{fig:art-fad} presents as several bands across the reconstructed dark-field images.      

\bibliographystyle{plain}
\bibliography{references}